\documentclass[a4paper,11pt]{article}

\usepackage{jcappub} 

\usepackage[T1]{fontenc} 

\title{Cosmogenic radionuclide production in NaI(Tl) crystals}

\author{J.~Amar\'e,}
\author{S.~Cebri\'an,}
\author[1]{C.~Cuesta,\note{Present address: Center for Experimental Nuclear Physics and Astrophysics, and Department of Physics, University of Washington, Seattle, WA, USA}}
\author{E.~Garc\'ia,}
\author{C.~Ginestra,}
\author{M.~Mart\'inez,}
\author{M.A.~Oliv\'an,}
\author{Y.~Ortigoza,}
\author{A.~Ortiz de Sol\'orzano,}
\author[2]{C.~Pobes,\note{Present address: Instituto de Ciencia de Materiales de Arag\'on, Universidad de Zaragoza - CSIC}}
\author{J.~Puimed\'on,}
\author{M.L.~Sarsa,}
\author{J.A.~Villar}
\author{and P.~Villar}
\affiliation{Laboratorio de F\'isica Nuclear y Astropart\'iculas, Universidad de Zaragoza
\\ Calle Pedro Cerbuna 12, 50009 Zaragoza, Spain
\\ Laboratorio Subterr\'aneo de Canfranc
\\ Paseo de los Ayerbe s/n, 22880 Canfranc Estaci\'on, Huesca,
Spain}

\emailAdd{amare@unizar.es} \emailAdd{scebrian@unizar.es}
\emailAdd{ccuesta@unizar.es} \emailAdd{edgarcia@unizar.es}
\emailAdd{cginestra@unizar.es} \emailAdd{mariam@unizar.es}
\emailAdd{maolivan@unizar.es} \emailAdd{ortigoza@unizar.es}
\emailAdd{alfortiz@unizar.es} \emailAdd{cpobes@unizar.es}
\emailAdd{puimedon@unizar.es} \emailAdd{mlsarsa@unizar.es}
\emailAdd{villar@unizar.es} \emailAdd{pvillar@unizar.es}

\abstract{The production of long-lived radioactive isotopes in
materials due to the exposure to cosmic rays on Earth surface can be
an hazard for experiments demanding ultra-low background conditions,
typically performed deep underground. Production rates of cosmogenic
isotopes in all the materials present in the experimental set-up, as
well as the corresponding cosmic rays exposure history, must be both
well known in order to assess the relevance of this effect in the
achievable sensitivity of a given experiment. Although NaI(Tl)
scintillators are being used in experiments aiming at the direct
detection of dark matter since the first nineties of the last
century, very few data about cosmogenic isotopes production rates
have been published up to date. In this work we present data from
two 12.5~kg NaI(Tl) detectors, developed in the frame of the ANAIS
project, which were installed inside a convenient shielding at the
Canfranc Underground Laboratory just after finishing surface
exposure to cosmic rays. The very fast start of data taking allowed
to identify and quantify isotopes with half-lives of the order of
tens of days. Initial activities underground have been measured and
then production rates at sea level have been estimated following the
history of detectors; values of about a few tens of nuclei per kg
and day for Te isotopes and $^{22}$Na and of a few hundreds for I
isotopes have been found. These are the first direct estimates of
production rates of cosmogenic nuclides in NaI crystals. A
comparison of the so deduced rates with calculations using typical
cosmic neutron flux at sea level and a carefully selected
description of excitation functions will be also presented together
with an estimate of the corresponding contribution to the background
at low and high energies, which can be relevant for experiments
aiming at rare events searches.}


\begin{document}
\maketitle \flushbottom

\section{Introduction}
\label{intro}

Production of radioactive isotopes by exposure to cosmic rays is a
well known issue \cite{lal}. Several cosmogenic isotopes are
important because of their broad applications in extraterrestrial
and terrestrial studies dealing with astrophysics, geophysics,
paleontology and archaeology. For experiments searching for rare
phenomena like the nuclear Double Beta Decay or the interaction of
Weakly Interacting Massive Particles (the so-called WIMPs, which
could be filling the galactic dark matter halo) requiring detectors
working in ultra-low background conditions, long-lived radioactive
nuclei induced in materials of the set-up can be a sensitivity
limiting hazard. Operating in deep underground locations, using
active and passive shields and selecting carefully radiopure
materials for the building of the detectors and shields are common
practices in this kind of experiments to reduce very efficiently the
background entangling the searched signal \cite{heusser,formaggio}.
Cosmogenic radionuclides produced during fabrication, transport and
storage of components may be even more important than residual
contamination from primordial nuclides and become very problematic,
depending on the target. For instance, the poor knowledge of cosmic
ray activation in detector materials is highlighted in
\cite{gondolo} as one of the three main uncertain nuclear physics
aspects of relevance in the direct detection approach pursued to
solve the dark matter problem.

One of the most relevant processes in cosmogenic activation is the
spallation of nuclei by high energy nucleons, but other reactions
like fragmentation, induced fission or capture can be very important
for some nuclei. Cosmogenic activation is strongly dependent on the
nucleon flux, neutron to proton ratio, and energies available. At
sea level, for instance, the flux of neutrons and protons is
virtually the same at a few GeV, but the proton to neutron ratio
decreases significantly at lower energies because of the charged
particles absorption in the atmosphere. For example, at 100~MeV this
ratio is about 3\% \cite{lal}. Summarizing, at sea level, nuclide
production is mainly dominated by neutrons at low energies, whereas
if materials are flown at high altitude cosmic flux is much greater,
energies at play are larger and activation by protons can not be
neglected.

In principle, cosmogenic activation can be kept under control by
minimizing exposure at surface and storing materials underground,
avoiding flights and even using shields against the hadronic
component of cosmic rays during surface operation. But since these
requirements usually complicate the preparation of experiments (for
example, while crystal growth and detectors mounting steps) it would
be desirable to have reliable tools to quantify the real danger of
exposing the different materials to cosmic rays. Direct
measurements, by screening of exposed materials in very low
background conditions as those achieved in underground laboratories,
and calculations of production rates and yields, following different
approaches, have been made for several materials in the context of
dark matter, double beta decay and neutrino experiments (see for
instance \cite{cebrian} and references therein). Many different
studies are available for germanium \cite{avignone}-\cite{cebrianap}
and interesting results have been derived also for copper
\cite{cebrianap,copper}, stainless steel \cite{steel}, tellurium
\cite{te,telozza} or neodymium \cite{nd1,nd2}. However, systematic
studies for other targets are missed.

Sodium iodide crystals (Tl doped) have been widely used as radiation
detectors profiting from the very high light output
\cite{birks,lecoq}. In particular, NaI(Tl) detectors have been
applied in the direct search for dark matter for a long time
\cite{fushimi}-\cite{fushimi2}. Among the several experimental
approaches using NaI(Tl) detectors, DAMA/LIBRA is the most relevant,
having reported the observation of a modulation compatible with that
expected for galactic halo WIMPs with a large statistical
significance \cite{dama}. Results obtained with other target
materials and detection techniques have been ruling out for years
the most plausible compatibility scenarios. The ANAIS (Annual
modulation with NaI Scintillators) project \cite{anais25taup} is
intended to search for dark matter annual modulation with ultrapure
NaI(Tl) scintillators at the Canfranc Underground Laboratory (LSC)
in Spain; the aim is to provide a model-independent confirmation of
the annual modulation positive signal reported by DAMA/LIBRA using
the same target and technique. Several prototypes have been
developed and operated in Canfranc \cite{anaisbkg}-\cite{analysis}.
Projects like DM-Ice \cite{dmice}, KIMS \cite{kims} and SABRE
\cite{sabre} also envisage the use of large masses of NaI(Tl) for
dark matter searches. However, complete understanding of the
DAMA/LIBRA background at low energy has not yet been achieved and
some open questions remain \cite{kudry,response}; therefore, careful
analysis of the different background components is highly
recommended. In the case of ANAIS, a successful background model was
built for previous prototypes (ANAIS-0) \cite{anaisbkg}, but
ANAIS-25 background is not completely explained by a similar
background model \cite{anais25}. Hence, other possible background
contributions have to be included into the model, and in particular,
we address in this work the cosmogenic contribution.

For NaI, production of $^{129}$I, $^{125}$I and $^{22}$Na was
analyzed in \cite{damanima2008} in the context of the DAMA/LIBRA
experiment and the presence of cosmogenic isotopes has been also
reported by ANAIS \cite{anais25,ichep}, DM-Ice \cite{dmice} and KIMS
\cite{kims} collaborations. In this work, cosmogenic yields induced
on the Earth surface in two large NaI(Tl) crystals produced by Alpha
Spectra\footnote{Alpha Spectra Inc., Grand Junction, Colorado, US.
\url{http://www.alphaspectra.com/}.} for the ANAIS experiment have
been quantified; results presented here come from the so-called
ANAIS-25 set-up \cite{anais25}. The ultra-low background conditions
of ANAIS and a prompt start of data taking after moving the
detectors underground were essential to carry out the analysis of
the cosmogenic activation. The structure of the paper is the
following. Section~\ref{mea} describes the experimental set-up and
the measurements taken, while the data analysis performed and the
production rates at sea level derived for the cosmogenic isotopes
identified in the measurements are presented in section~\ref{ana}. A
calculation of the corresponding production rates has been attempted
using carefully selected excitation functions and the history of the
raw material the detectors consist of; such calculations are
summarized in section~\ref{cal}, together with the comparison with
the experimental rates and corresponding discussion. The impact of cosmogenic products on the background levels at the energy regions relevant for rare
event searches is evaluated in section~\ref{contri}. Finally,
conclusions are presented in section~\ref{con}.

\section{Experimental set-up and measurements}
\label{mea}

\begin{figure}
 \centering
 \includegraphics[width=10cm]{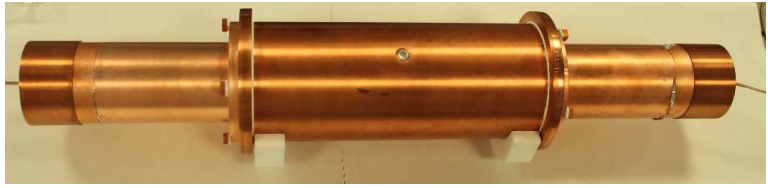} 
 \caption{Picture of one of the ANAIS-25 detectors (D0), after the coupling of PMTs and placement of the copper casing at the LSC clean room.}
  \label{setup}
\end{figure}

\begin{figure}
 \centering
 \includegraphics[width=8cm]{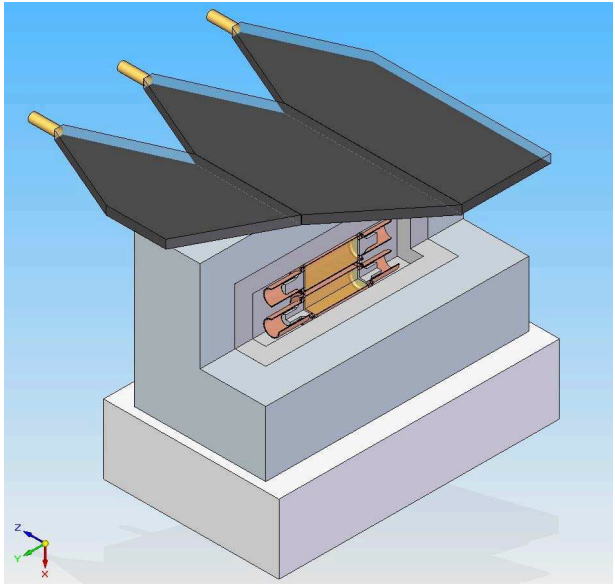}
 \caption{Schematic drawing of the ANAIS-25 experimental layout at LSC, consisting of 10~cm archaeological lead plus 20~cm low activity lead, enclosed in a PVC box continuously flushed with boil-off nitrogen and active vetoes anti-muons.}
  \label{setup2}
\end{figure}

Two NaI(Tl) detectors built in collaboration with Alpha Spectra form
the ANAIS-25 set-up \cite{anais25}.
Each detector consists of a cylindrical 12.5~kg NaI(Tl) crystal,
with 4.75'' diameter and 11.75'' length. They were grown with
selected ultrapure NaI powder and
encapsulated in OFHC copper with two synthetic quartz windows
allowing the coupling to photomultiplier tubes (PMTs) in a second
step. Two Hamamatsu R12669SEL2 PMTs were used for one crystal and
Hamamatsu R11065SEL PMTs for the other; we will refer to both
detectors as D0 and D1, respectively, in the following. Only white
Teflon was used as light diffuser, wrapping the crystal, inside the
copper encapsulation. A Mylar window allows to calibrate at low
energy both detectors.
ANAIS-25 detectors are being operated at hall B of LSC under 2450
m.w.e. (meter water equivalent), inside a shielding consisting of
10~cm archaeological lead plus 20~cm low activity lead, all enclosed
in a PVC box tightly closed and continuously flushed with boil-off
nitrogen. Figure~\ref{setup} shows a picture of detector D0 and a
drawing of the ANAIS-25 set-up is depicted in figure~\ref{setup2}.

Each PMT charge output signal is separately processed
\cite{anais25}; each one is divided into a trigger signal, a signal
going to the digitizer, and several ones differently
amplified/attenuated, and fed into QDC (charge-to-digital converter)
module channels to be integrated in a 1~$\mu$s window. Low energy
(LE) and high energy (HE) ranges, below and above $\sim$200~keV
respectively, have been considered in the data analysis presented
here. The building of the spectra was done by software (off-line) by
adding the signals from both PMTs. Data triggering is done in
logical OR mode between the two detectors and in logical AND between
the two PMT signals from each detector, at photoelectron level. The
outstanding light collection efficiency obtained in ANAIS-25
modules, with measured values of 16.13$\pm$0.66 and
12.58$\pm$0.13~phe/keV for D0 and D1 detectors respectively
\cite{anais25} guarantees a very low energy threshold.

The main goal of ANAIS-25 set-up was to assess the total background
of the detectors and, in particular, to determine precisely the
$^{40}$K content of the crystals, since it was found to be the
dominant background in previous ANAIS prototypes \cite{anaisbkg}.
The technique used to estimate the bulk $^{40}$K activity in the
crystals is the measurement in coincidence: one detector measures
the total energy (3.2~keV) released by the X-ray/Auger electrons
emissions of argon following the K-shell electron capture (EC) of
$^{40}$K, while the other one measures the fully absorbed energy
(1460.8~keV) of the high energy gamma escaping from the former.
Averaging the results for the two crystals, an activity of
1.25$\pm$0.11~mBq/kg (41.7$\pm$3.7~ppb of natural potassium) was
derived \cite{ijmpa}; this value is one order of magnitude better
than that found in other crystals, but still a factor of 2 higher
than the required purity. Background assessment is in progress, as
far as cosmogenic contribution to the background is still decaying.
The excellent light collection efficiency measured in ANAIS-25
modules results in a very good energy resolution which has helped in
the identification of cosmogenically produced isotopes.

ANAIS-25 crystals were built in Alpha Spectra facilities in Colorado
and then shipped by boat from the US to Spain, avoiding air travel.
They arrived at LSC in the evening of 27$^{th}$ November 2012 and
were immediately stored underground; data taking started only three
days later, after coupling PMTs in the LSC clean room, placing the
detectors inside the shielding and preliminary testing of electrical
connections and general performance. This prompt commissioning of
ANAIS-25 allowed to observe short-lived isotopes and to properly
quantify induced activation in those long-lived. Data analyzed for
this study correspond to three data sets. Data sets I and II span
altogether for 210~days, from the beginning of the data taking to
the end of June 2013; we will refer to set I for data obtained until
February 2013 and to set II for those taken from March to June 2013.
They were acquired in slightly different gain conditions, which
implied that some signals present in the LE or HE spectra in one set
were not available in the other. Set III includes 88~days of data
from March to June 2014, once the contribution of most of the
cosmogenic nuclei is significantly reduced, allowing to identify
longer-living products.

\begin{figure}
 \centering
 \includegraphics[width=8cm]{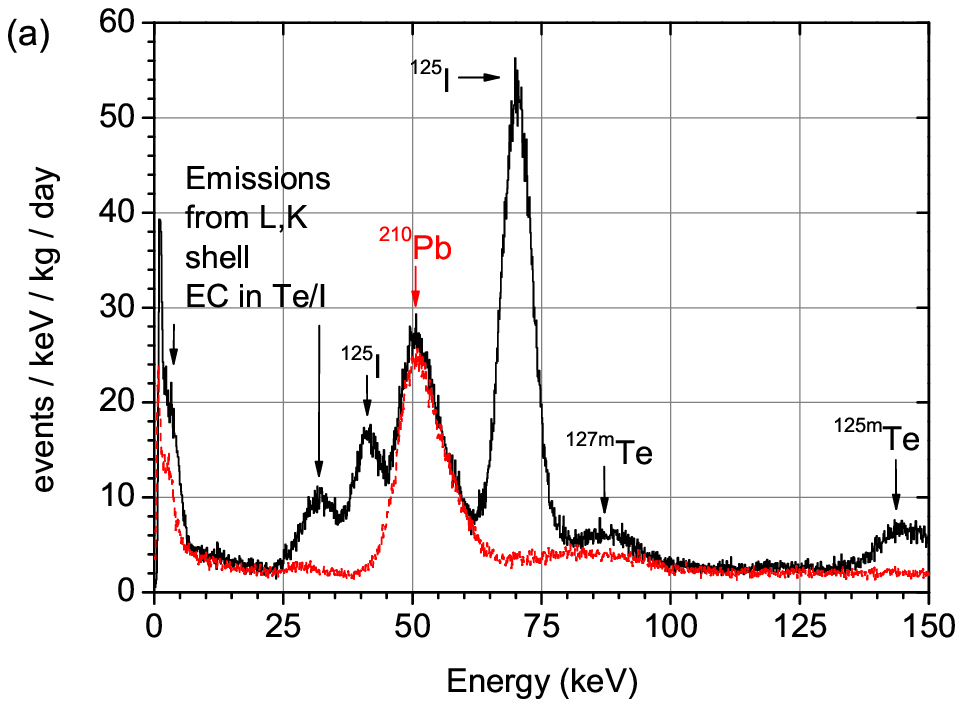}
 \includegraphics[width=8cm]{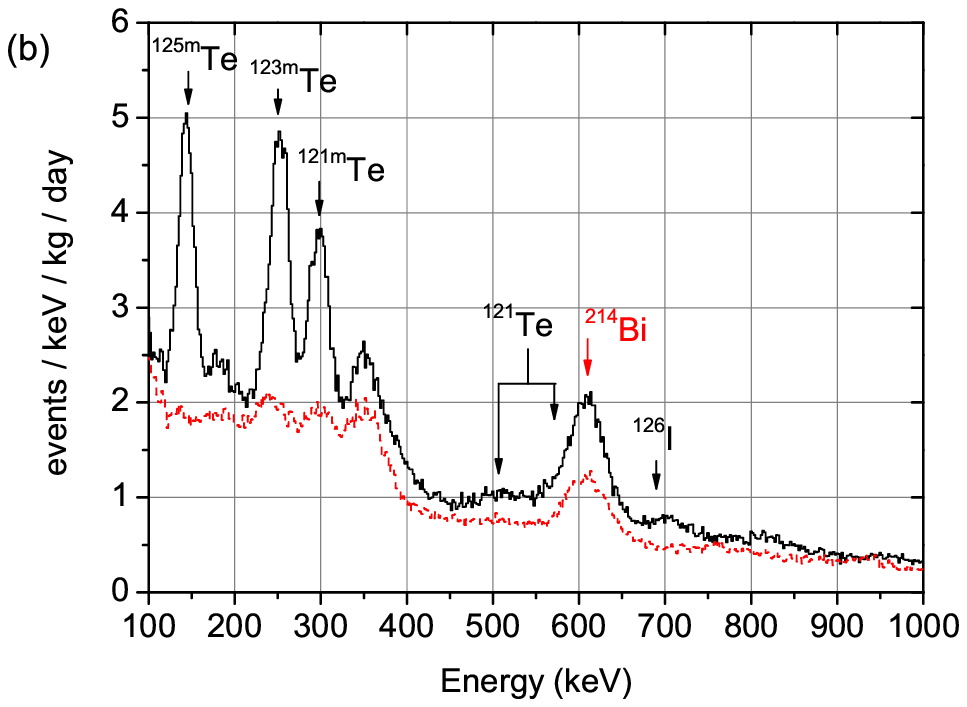}
 \caption{Comparison of the ANAIS-25 spectra evaluated in the beginning of data set I (black solid line) and about fifteen months afterwards, during data set III (red dashed line), over 13.0 and 18.5~days,
respectively. The low (a) and high (b) energy regions are shown.
Only data from D0 are shown, being those for D1 very similar.
Several cosmogenic emissions have been identified and are labeled
(see text and table~\ref{isotopes}). Main background lines are
labeled too (in red).}
  \label{difference}
\end{figure}

Figure~\ref{difference} compares the spectra from D0, both in the
low and high energy regions, evaluated in the beginning of data set
I and about fifteen months afterwards, during data set III, over
13.0 and 18.5~days, respectively. Several lines can be clearly
attributed to cosmogenic activation, as figure~\ref{difference}
shows. Iodine and tellurium isotopes identified are presented in
table~\ref{isotopes}; recommended values for half-lives by Decay
Data Evaluation Projet (DDEP)
\cite{DDEP}\footnote{\url{http://www.nucleide.org/DDEP_WG/Periodes
2010.pdf}} have been considered for all the analyzed nuclides.
Although not directly seen in figure~\ref{difference},
cosmogenically produced $^{22}$Na, having a longer mean life than I
and Te products, has been identified and quantified using
coincidence spectra along data set III.

\begin{table}
\centering
\begin{tabular}{r@{}lcccc}
\hline \multicolumn{2}{c}{Isotope} &  Half-life & Decay mechanism &
Main $\gamma$ emissions /  & $I\epsilon$\\
&  & & & Metastable level energy (keV) & \\ \hline

$^{126}$&I & (12.93$\pm$0.05) d  & EC, $\beta^{+}$, $\beta^{-}$ & 666.3 & 0.0035\\ 

$^{125}$&I &  (59.407$\pm$0.009) d  &EC &35.5 & 0.8011\\ 

$^{127m}$&Te&  (107$\pm$4) d  &IT, $\beta^{-}$ & 88.3 & 0.9727\\  

$^{125m}$&Te& (57.4$\pm$0.2) d   &IT& 144.8 & 0.968 \\ 

$^{123m}$&Te & (119.3$\pm$0.1) d  &IT &247.6 & 0.972\\ 

$^{121m}$&Te & (154$\pm$7) d  &IT, EC& 294.0 & 0.842\\ 

$^{121}$&Te & (19.16$\pm$0.05) d &EC &507.6, 573.1 & 0.011 \\ 

$^{22}$&Na & (2.6029$\pm$0.0008) y & EC, $\beta^{+}$ & 511 & 0.0050
\\ 

& &   & & 1274.6 & 0.0038 \\ 

\hline

\end{tabular}
\caption{\label{isotopes} Main cosmogenically induced isotopes
identified in the data of ANAIS-25 modules. Half-lives and decay
mechanisms are indicated. Fourth column shows the energy of main
gamma emissions or, in the case of the metastable states, the energy
of the corresponding excited level. Last column gives the considered
product of intensity I and detection efficiency $\epsilon$ for the
analyzed distinctive signal of each isotope (see text). Decay
information is taken from \cite{DDEP} for $^{125}$I, $^{127m}$Te,
$^{123m}$Te and $^{22}$Na and from \cite{toi} for the other
isotopes.}
\end{table}

There are other isotopes, also cosmogenically produced, that have
not been identified in this analysis, although they are probably
present in ANAIS-25 crystals. It is worth to mention the cases of
$^{129}$I and $^{123}$Te.

$^{129}$I can be produced by uranium spontaneous fission and by
cosmic rays. Its concentration is strongly affected by the ore
material exposure either to cosmic rays or to high uranium content
environment. It presents 100\% $\beta^{-}$ decay to the excited
level of 39.6~keV of the daughter nucleus with a half-life
$T_{1/2}=(16.1\pm0.7)10^{6}$ y \cite{DDEP}, being hence the expected
signal in large NaI(Tl) crystals a continuous beta spectrum starting
in 39.6~keV. This signal is above the region of interest for dark
matter searches, but it is important for background understanding
\cite{anaisbkg}. The long lifetime and the difficulty to disentangle
the signal from other emissions are the reasons why we cannot
quantify the amount of $^{129}$I in ANAIS-25 crystals. In
\cite{damanima2008} the estimated fraction of this isotope was
determined to be $^{129}I/^{nat}I=(1.7\pm0.1)10^{-13}$; however,
strong variability of this concentration is expected in different
origin ores.

In the case of $^{123}$Te, we expect the presence of this isotope as
daughter of the metastable state $^{123m}$Te, effectively
identified, and probably also directly produced by cosmic rays by
similar reaction mechanisms. This isotope would decay 100\%
following EC to the ground state of the daughter nucleus, having a
half-life larger than 10$^{13}$ y \cite{toi}. The signature in
ANAIS-25 detectors would be a peak corresponding to the full
absorption of the binding energy of L-shells of Sb. For the moment,
this signal cannot be resolved from other contributions from Te/I
isotopes still present, having L-shell EC in higher or lower extent,
and then quantification is not possible in ANAIS-25 data.

\section{Data analysis and estimate of production rates at sea level}
\label{ana}

We analyze in the following the distinctive signatures considered
for the identification of each cosmogenically induced isotope listed
in table~\ref{isotopes}, as well as the estimate of the
corresponding detection efficiency $\epsilon$ in ANAIS-25 set-up.
When required, the efficiencies have been obtained by Geant4
simulation (version 9.4.p01) considering the complete geometry of
the ANAIS-25 set-up and isotopes homogeneously distributed in the
crystals; this kind of Monte Carlo simulations was validated with a
similar set-up, but slightly different detector geometry, in
\cite{anaisbkg} by reproducing measurements with calibration
sources, finding an accuracy at the level of 10\%, although a
Compton/peak suppression is systematically observed in the
simulations. For the metastable isotopes of Te, full energy
depositions corresponding to the excited level energy have been
considered.
Activation has been assessed independently for each
crystal, and whenever possible, results have been properly combined.

\begin{figure}
 \centering
 \includegraphics[width=8cm]{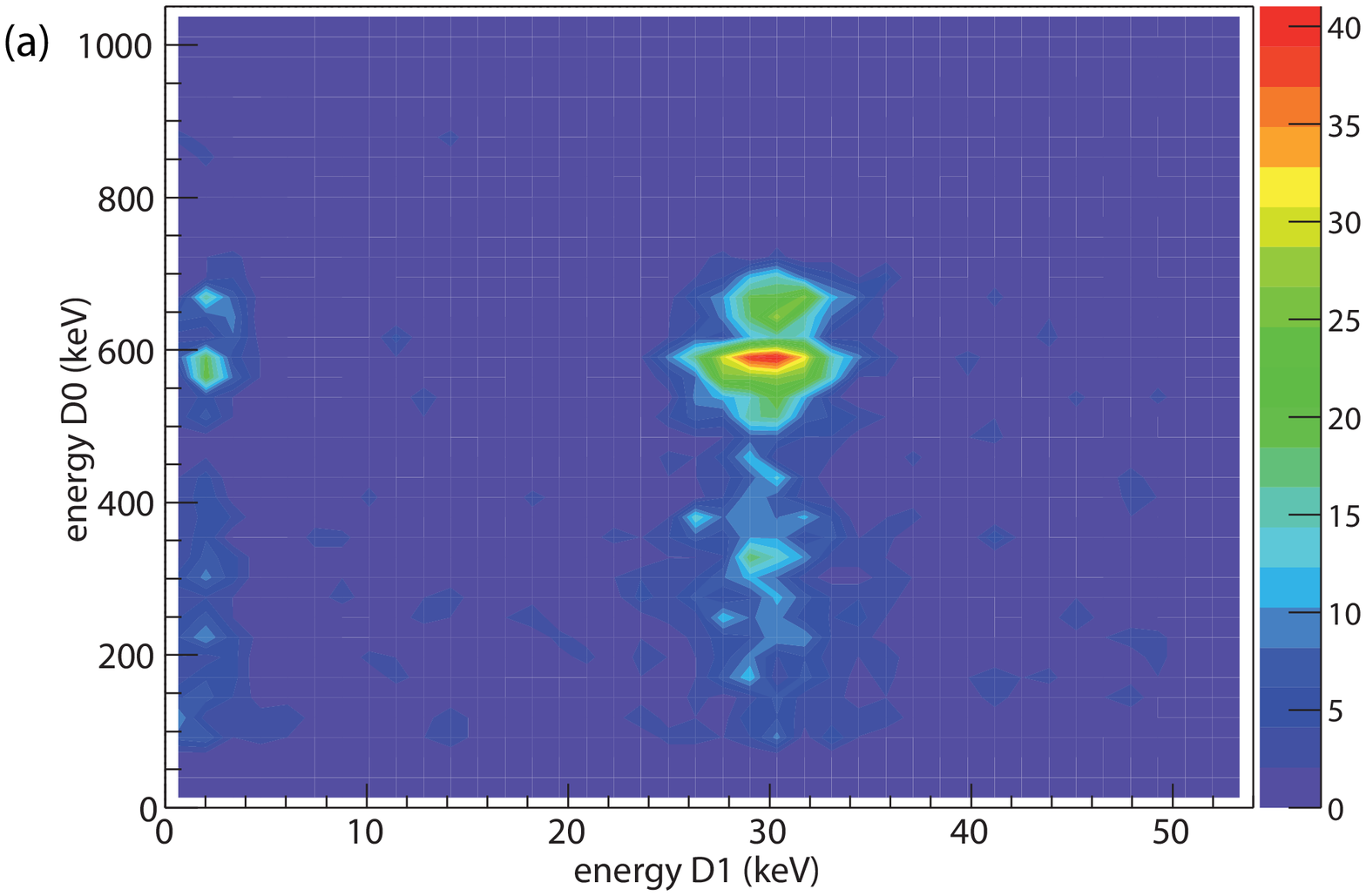}
 \includegraphics[width=8cm]{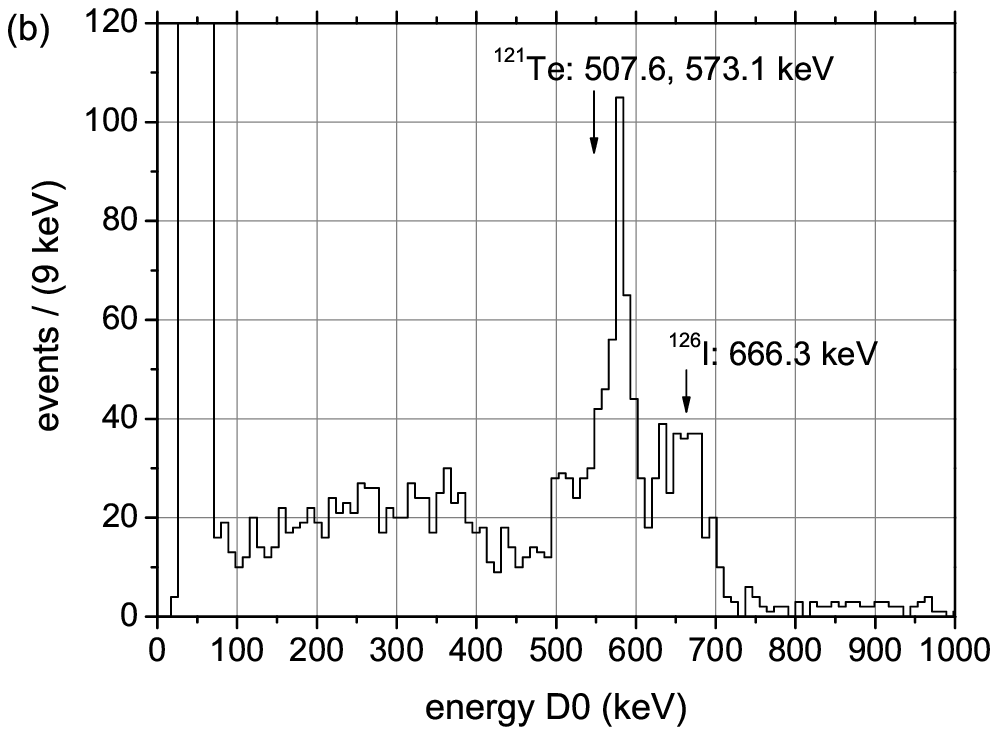}
 \caption{(a): Coincidence plot between D0 and D1 in the beginning of data set I. Coincidences due to both $^{126}$I and $^{121}$Te between high energy photons absorbed in D0 and the binding energies of Te or Sb registered in D1 are clearly seen, for K shell (around 31~keV) and even for L shell (around
 5~keV). D0 energies below 80 keV have been cut because the high rate of coincidences
 there prevents from seeing $^{126}$I and $^{121}$Te coincidences. (b): Spectrum of the high energy D0 data in coincidence with a window in D1 data around K shell binding energies of Te and Sb; peaks at 507.6-573.1 and 666.3~keV are singled out and can be quantified.}
  \label{coincidences}
\end{figure}

\begin{itemize}
\item For $^{125}$I, the peak at 67.3~keV has been considered. Since this isotope decays with 100\% probability by electron capture to an excited state of the Te daughter nucleus,
this peak is produced by the sum of the energy of the excited level
(35.5~keV) and the binding energy of the K shell of Te (31.8~keV).
Efficiency for the detection of the energy deposition is expected to
be $\sim$100\%, hence the product $I\epsilon$ is taken to be equal
to the K-shell EC probability, $I\epsilon=$0.8011 \cite{DDEP}.
\item For $^{127m}$Te, the peak at 88.3~keV, corresponding to the energy of the metastable state, has been
analyzed. The branching ratio of the internal transition is
$I=$97.27\% \cite{DDEP} and 100\% detection efficiency has been
assumed. Signal from D1 detector could not be evaluated along data
set II since a worse resolution hindered good fitting and therefore
in this case final results come only from D0 detector.
\item For $^{125m}$Te, the peak at 144.8~keV has been studied. Internal
transition occurs with 100\% probability and detection efficiency
has been evaluated by simulation giving $\epsilon=$0.968. In this
case, it has been possible to quantify the peak from both the low
and high energy spectra, although not for all data sets and
detectors; therefore, final results have been obtained only from D0
in the low energy spectrum.
\item For $^{123m}$Te, the peak at 247.6~keV in high energy spectra has been
analyzed. The branching ratio of internal transition is 100\%. The
loss of efficiency due to the emission with 84\% probability of a
159~keV photon \cite{DDEP} which can escape detection has been
evaluated by Geant4 simulation giving an overall detection
efficiency $I\epsilon=$0.972.
\item Similarly, for $^{121m}$Te, the peak at 294.0~keV in high energy spectra has been
analyzed and a correction factor to the detection efficiency has
been estimated by means of Geant4 simulation to account for the
probability of losing the 212~keV gamma ray which is emitted in 81\%
of the decays \cite{toi}. The branching ratio of internal transition
is here 88.6\% \cite{toi} and then the final value considered for
$I\epsilon$ is 0.842.
\item A coincidence technique analysis has been applied to quantify $^{126}$I
and $^{121}$Te, which decay by electron capture to excited states of
the daughter nuclei. The identifying signature is the number of
coincidences produced by the full absorption of high energy gamma
emissions (see table~\ref{isotopes}) in one detector and that
corresponding to the binding energy for K shell electrons in Te or
Sb (31.8 or 30.5~keV, respectively) in the other detector. Just as
an example, figure~\ref{coincidences} shows the coincidence plot
between detectors D0 and D1, considering their high and low energy
spectra respectively, obtained in the beginning of data set I. 
By selecting a window from 15 to 45~keV in D1 data, we obtain the D0
coincident spectrum shown in figure~\ref{coincidences}: peaks from
$^{121}$Te and $^{126}$I are clearly observed and are used to
determine the number of events attributable to each isotope in a
given time period. The product $I\epsilon$ for these coincidence
signals has been evaluated by simulating the decays of each isotope
in the crystals with Geant4, obtaining $I\epsilon=$0.0035 for
$^{126}$I and $I\epsilon=$0.011 for $^{121}$Te.
These two isotopes have the shortest half-lives of all the analyzed
products. The coincidence signal has been properly identified only
within the first 42~days of available data for $^{126}$I. However,
for $^{121}$Te the signal has not faded out since this isotope is
being produced as result of the decay of the longer lived
$^{121m}$Te, also cosmogenically induced; therefore, this point has
been considered in the analysis of the time evolution of the
coincidence signature of $^{121}$Te (see
section~\ref{subsectionte121}).
\item Presence of $^{22}$Na, having a longer lifetime, is more subtle. The analyzed signal has been the integral number of events from 2300 to 2900~keV (corresponding to full absorption of its positron and gamma
emissions) in the spectrum obtained summing D0 and D1 energies, for
coincidence events leaving 511 (or 1275)~keV at any detector. This
is the best signature found for this isotope to avoid the
interference of backgrounds. The product $I\epsilon$ for these
signals has been evaluated by Geant4 simulation of $^{22}$Na decays
in the crystals, obtaining $I\epsilon=$0.0050 for the 511~keV peak
and $I\epsilon=$0.0038 for the 1275~keV signal.
\end{itemize}

In order to quantify the cosmogenic production of every isotope
identified in ANAIS-25 measurements, the cosmogenically induced
activity, which is the initial activity $A_{0}$ corresponding to the
moment of storing crystals deep underground at LSC, has been first
deduced (section~\ref{A0}). For that, the time evolution of the
identifying signature produced by each I and Te isotope has been
studied considering data sets I and II; as already commented,
special analysis is required for $^{121}$Te. For $^{22}$Na, the
integrated signal along data set III has been used. Finally, an
attempt has been made to estimate production rates $R_{p}$ at sea
level from those $A_{0}$ activities (section~\ref{Rp}), taking into
account the history of crystals.

\subsection{Initial activities underground}
\label{A0}

\subsubsection{I and metastable Te isotopes}
\label{ITe}

\begin{figure}
 \centering
 \includegraphics[width=7cm]{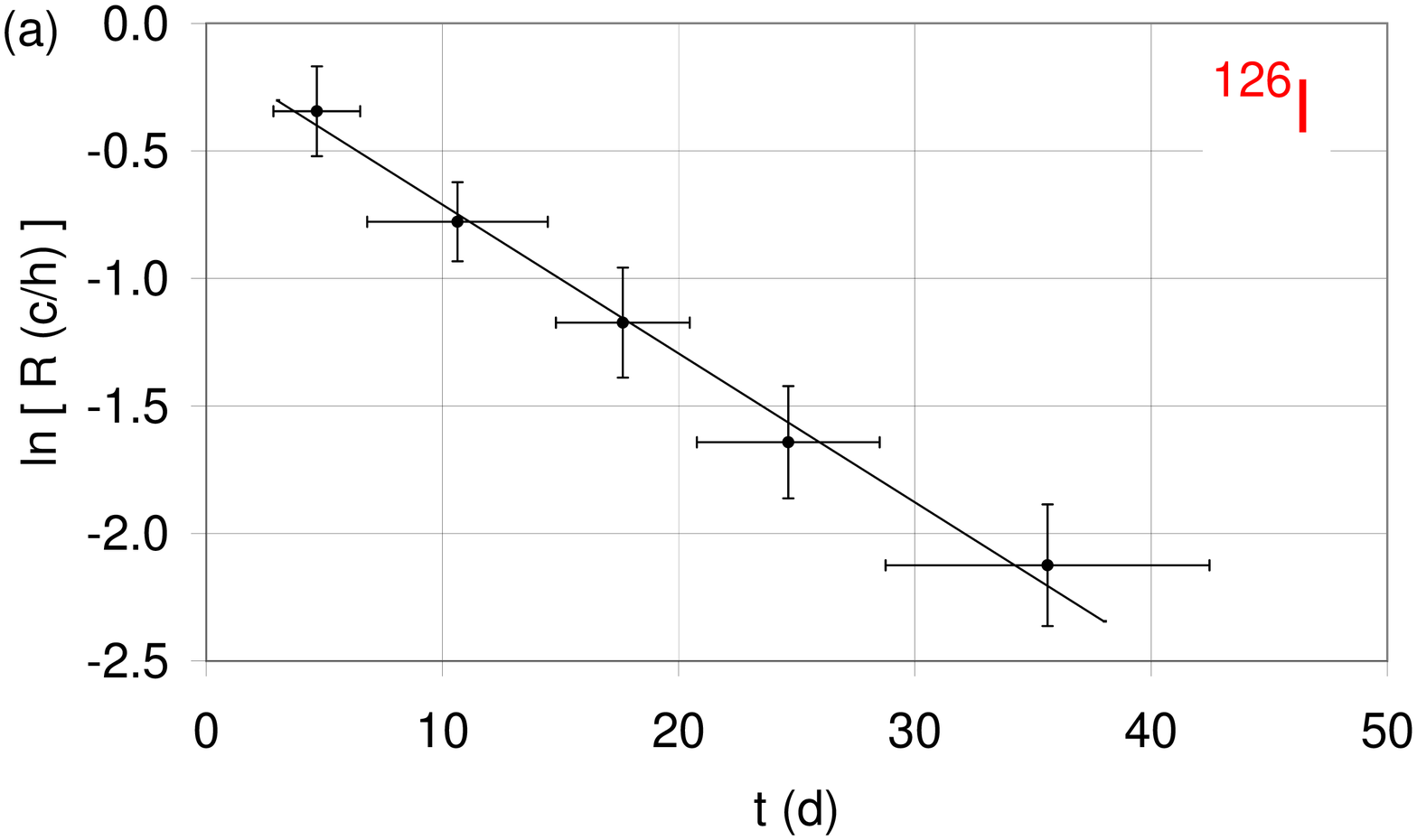} 
 \includegraphics[width=7cm]{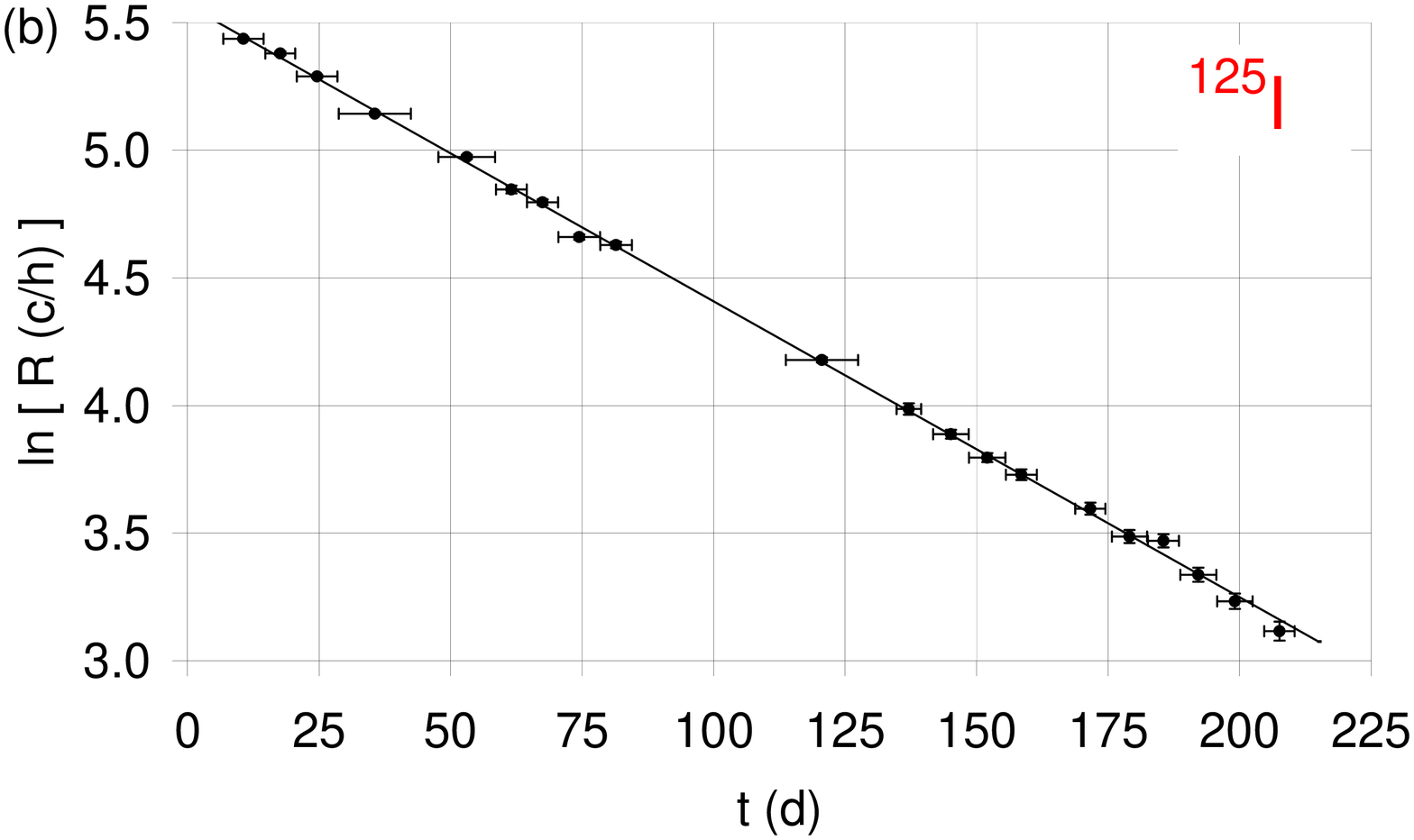}
 \includegraphics[width=7cm]{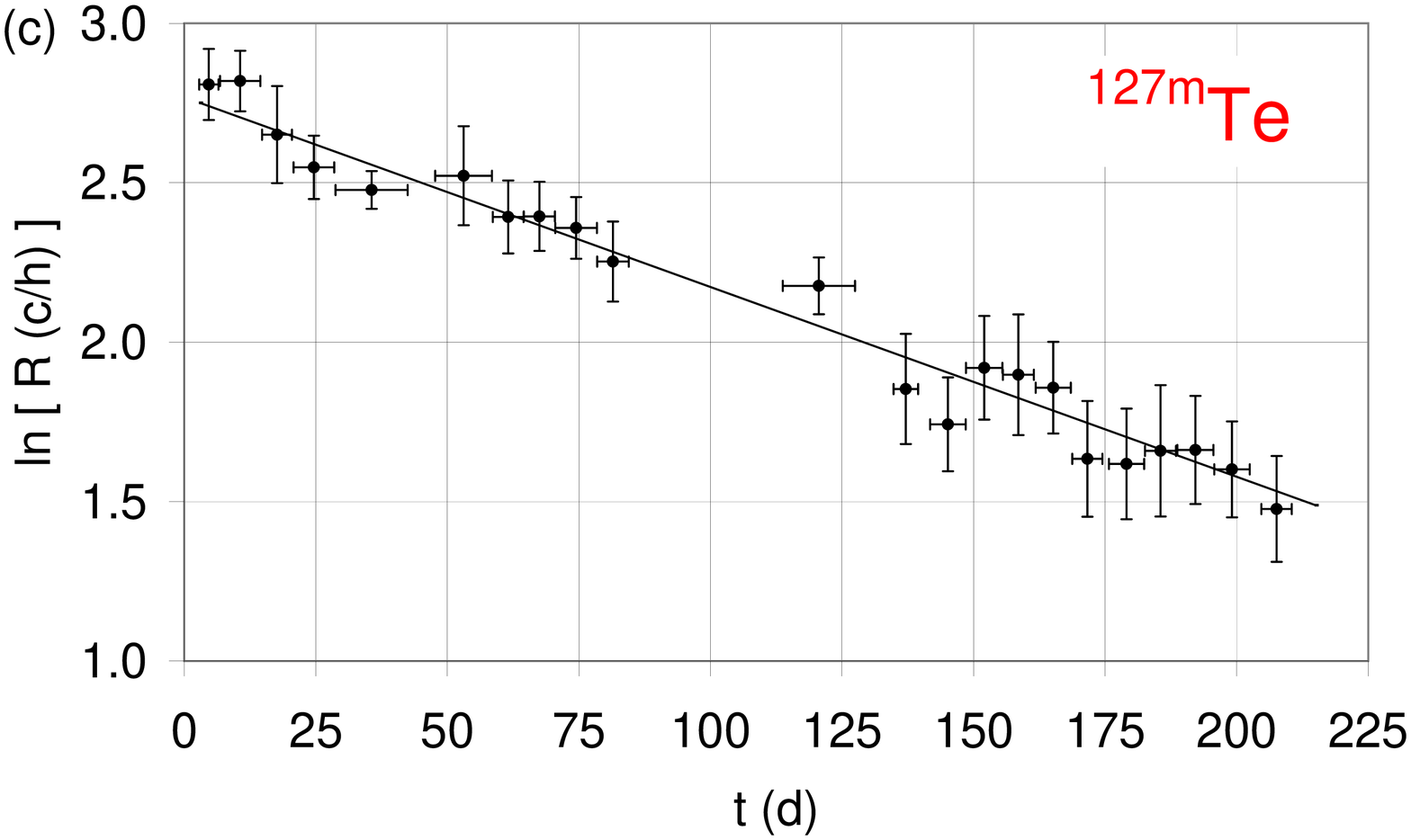}
 \includegraphics[width=7cm]{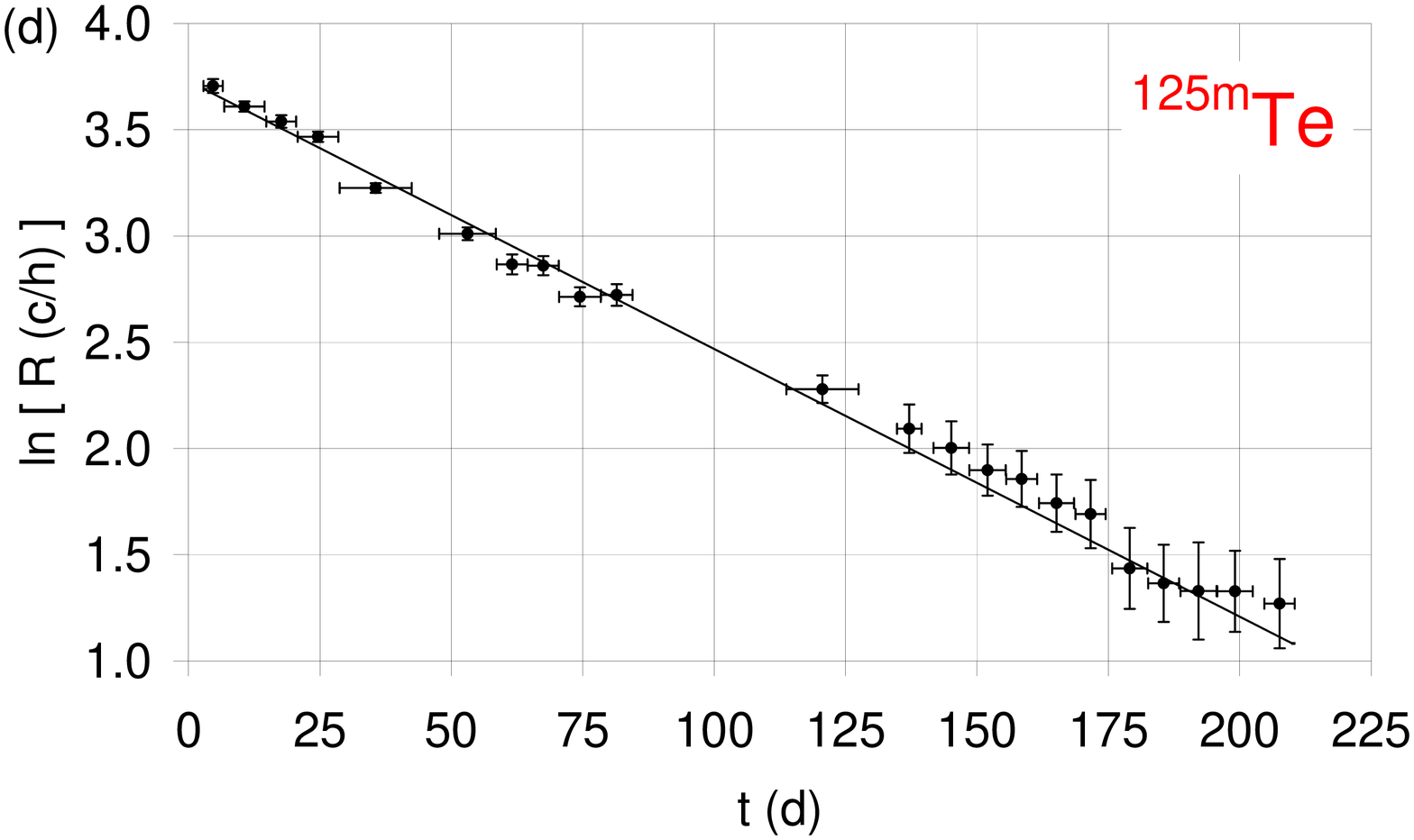}
 \includegraphics[width=7cm]{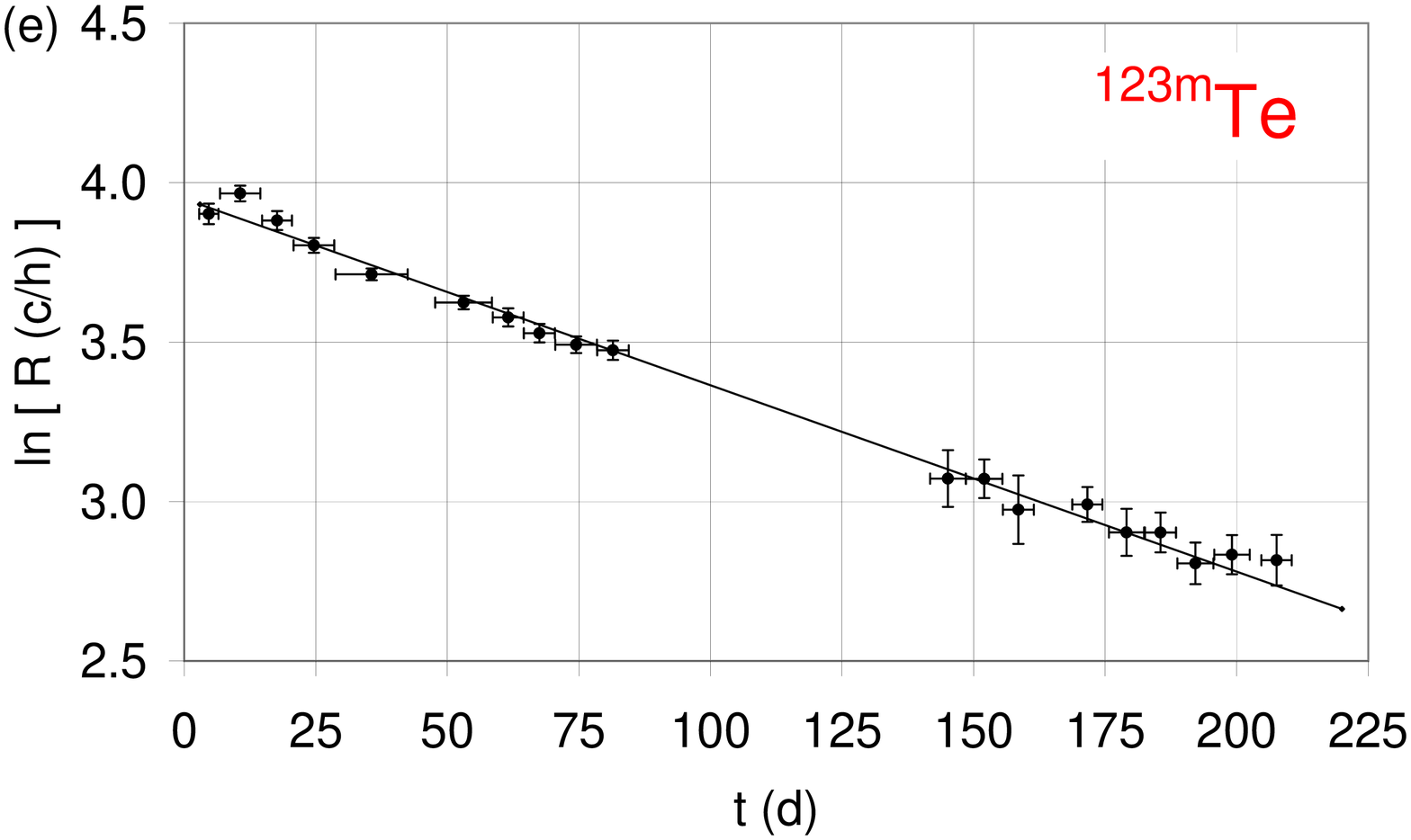}
 \includegraphics[width=7cm]{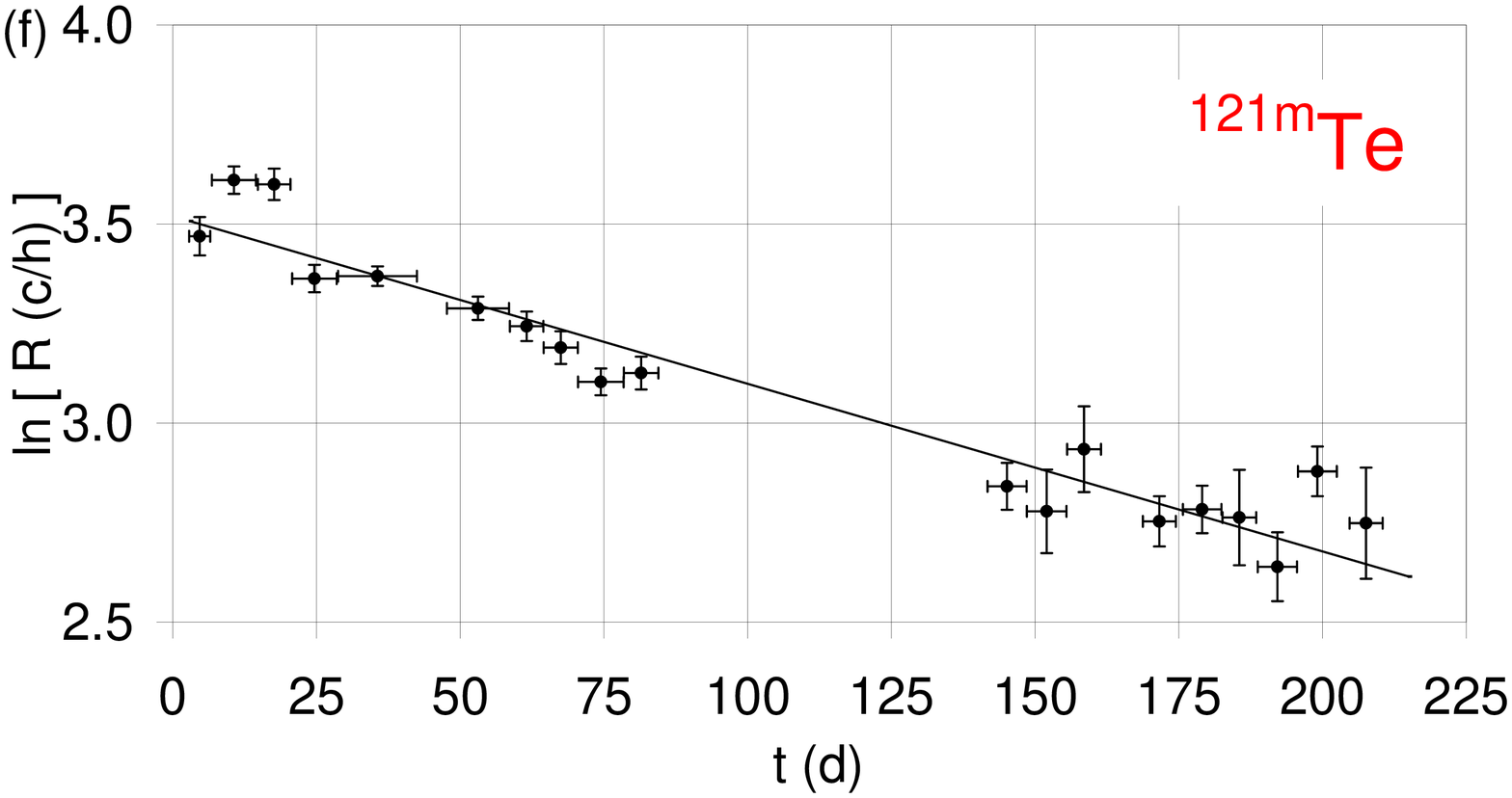}
 \caption{Evolution in time of the logarithm of the counting rate $R$ (expressed in counts per hour) assigned to each isotope,
 calculated as the addition of the rates in D0 and D1 detectors. Only for $^{127m}$Te (c) and $^{125m}$Te (d) results correspond to rate
 measured with D0 detector. The corresponding fits are also shown (solid line).
 Error bars on X-axis only indicate the time range signals have been evaluated over.}
  \label{plotsfits}
\end{figure}

The evolution in time of the counting rates $R$ for the different
identifying signatures described above has been analyzed and is
shown in the plots of figure~\ref{plotsfits} for I and metastable Te
isotopes\footnote{The origin of absolute times has been chosen at 0
h on 28$^{th}$ November 2012, corresponding to the storage of
detectors deep underground at LSC}. Single exponential decays
following the radioactive decay law are clearly observed in all of
them. Fits of $R$ vs time have been carried out considering not only
the sum of the signals from each detector (shown for most of the
products) but also independent signals from D0 and D1, and for
$^{125m}$Te, even signals from the low and high energy spectra have
been independently analyzed; in all cases compatible results have
been obtained. From the fit parameters the half-lives $T_{1/2}$ and
initial activities underground $A_{0}$ have been derived for each
isotope, taking into account the described intensity I and
efficiency $\epsilon$ for the considered signals, and are summarized
in table~\ref{finalfits}.
Error bars plotted in figure~\ref{plotsfits} represent for Y-axis
the uncertainty of peak areas, which has been taken into account in
the fits; for X-axis, they only indicate the time range signals have
been evaluated over.
As shown in table~\ref{finalfits},
the half-lives deduced from the fits agree with the known values
within uncertainties, confirming the choice of signatures for each
isotope.

The initial activities $A_{0}$ have been also deduced following a
different approach, taking into consideration the decay of isotopes
during the time intervals over which the identifying signals are
integrated. For an isotope with decay time constant $\lambda$, the
expected number of events $C$ from time $t_{i}$ to $t_{f}$ is:
\begin{equation}
C=\frac{A_{0} I \epsilon [\exp(-\lambda t_{i})-\exp(-\lambda
t_{f})]} {\lambda} \label{othermethod}
\end{equation}
An average of the $A_{0}$ values deduced from
eq.~(\ref{othermethod}) for each period of time has been obtained
for each isotope, weighted with uncertainties coming from peak area
estimates; results are shown in table~\ref{finalfits} too. The two
methods give compatible results, although slightly lower values are
systematically obtained from the second one; only a larger
difference is found for $^{126}$I due to its shorter half-life. The
advantage of deriving $A_{0}$ by fitting $R$ vs time is that no
assumption on the nature of the isotope is made, while the value
from the average of the initial activities obtained from time
intervals takes into account the decay of isotopes during those
intervals, which is not completely negligible.

\begin{table}
\centering
\begin{tabular}{lcccccc}
\hline Isotope & Known $T_{1/2}$  & Detector  &  Data sets & Fitted
$T_{1/2}$  & Fitted $A_{0}$  & Average $A_{0}$  \\ 
& (days)& & & (days) & (kg$^{-1}$d$^{-1}$) & (kg$^{-1}$d$^{-1}$) \\
\hline
$^{126}$I & 12.93$\pm$0.05  & D0+D1 &  I & 11.9$\pm$1.7 & 483$\pm$77 & 430$\pm$37 \\
$^{125}$I  & 59.407$\pm$0.009 & D0+D1 &   I+II& 59.80$\pm$0.26   & 627.6$\pm$2.4  & 621.8$\pm$1.6\\
$^{127m}$Te  & 107$\pm$4 & D0 &  I+II & 116.4$\pm$8.1 & 31.5$\pm$1.3 & 32.1$\pm$0.8\\
$^{125m}$Te  & 57.40$\pm$0.15 & D0, LE &  I+II &  55.0$\pm$1.1 & 82.6$\pm$1.2 & 79.1$\pm$0.8\\
$^{123m}$Te   & 119.3$\pm$0.1 & D0+D1 &  I+II & 118.6$\pm$3.1 & 102.5$\pm$1.2 & 100.8$\pm$0.8\\
$^{121m}$Te  & 154$\pm$7 & D0+D1 & I+II & 164.6$\pm$9.3 & 77.1$\pm$1.7 & 76.9$\pm$0.8\\
 \hline
\end{tabular}
\caption{\label{finalfits} Results derived for each cosmogenic
product from the selected detectors and data sets: half-lives
$T_{1/2}$ and initial activities underground $A_{0}$ from the fits
of $R$ vs time and average $A_{0}$ from activities obtained from
time intervals considering the isotope decay (see text). The known
values of the half-lives, presented in table~\ref{isotopes}, are
included again for comparison.}
\end{table}

\subsubsection{$^{121}$Te}
\label{subsectionte121}

\begin{figure}
 \centering
 \includegraphics[width=10cm]{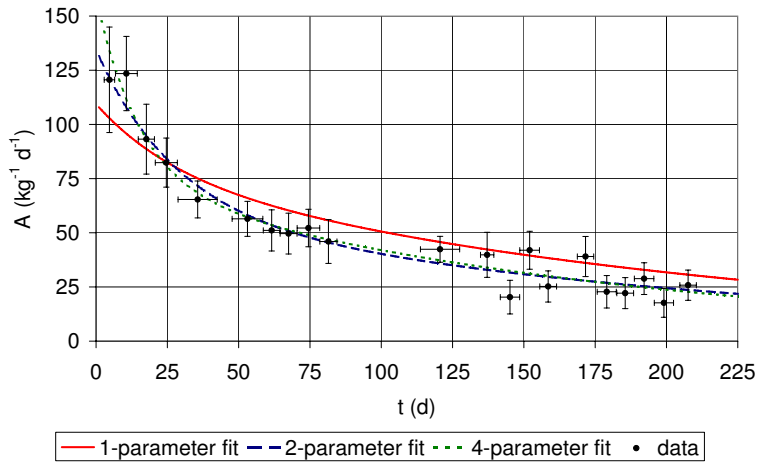}
 \caption{Evolution in time of the $^{121}$Te activity deduced from the identifying signature summed from D0 and D1;
 the lines show the corresponding fits to eq.~(\ref{ate121}), considering 1 (red solid line), 2 (blue dashed line) and 4 (green dotted line) free parameters (see text).}
  \label{fitte121}
\end{figure}

Figure~\ref{fitte121} shows the evolution in time of the activity of
$^{121}$Te deduced from its identifying signal adding D0 and D1
data. A single exponential decay with the half-life of $^{121}$Te is
not observed in this case because the isotope is being produced by
the decay of $^{121m}$Te, also cosmogenically induced and having a
longer half-life. A different fitting function is needed to derive
the initial activity underground and to estimate the production rate
at sea level for $^{121}$Te.

The time dependence of the activity of $^{121}$Te during the data
taking in ANAIS-25 can be written as the addition of two terms,
corresponding to the decay of the cosmogenically produced $^{121}$Te
and to the production, and subsequent decay, of $^{121}$Te from
$^{121m}$Te:
\begin{equation}
A(t) = \lambda_{g} N_{0,g} exp(-\lambda_{g}t) + P_{IT} N_{0,m}
\frac{\lambda_{g}\lambda_{m}}{\lambda_{g}-\lambda_{m}}
[\exp(-\lambda_{m}t)-\exp(-\lambda_{g}t)] \label{ate121}
\end{equation}
with subindex $g (m)$ referring to the ground (metastable) states of
the isotope and being $N_{0}$ the initial number of nuclei
(therefore those cosmogenically induced) and $P_{IT}=$0.886 the
probability of internal transition of the metastable state. To
obtain the initial number of nuclei, and then the corresponding
activities, for both the ground and metastable states, several fits
to the dependence given by eq.~(\ref{ate121}) for the signal from
$^{121}$Te have been tried and are also plotted in
figure~\ref{fitte121}. One fit has been performed letting four
parameters free: the activities $A_{0,g}=N_{0,g}\lambda_{g}$ and
$A_{0,m}=N_{0,m}\lambda_{m}$ and the half-lives $T_{1/2,g}$ and
$T_{1/2,m}$. Other fit has been made with two free parameters,
fixing the two half-lives according to the known values reported in
table~\ref{isotopes}. The last fit has been carried out with only
one free parameter, fixing in addition the activity of the
metastable state to the average of the two estimates given in
table~\ref{finalfits}: $A_{0,m}=$77.0~kg$^{-1}$d$^{-1}$. The product
$I\epsilon$ presented in table~\ref{isotopes} has been used. All the
fit results are presented in table~\ref{resultste121}. The three
estimates of the initial activity underground for $^{121}$Te are
reasonably compatible within uncertainties and those for $^{121m}$Te
are of the same order than the value deduced before (see
section~\ref{ITe} and table~\ref{finalfits}). The half-lives deduced
in the first fit reported in table~\ref{resultste121} are in the
2$\sigma$ interval of the known values. The $^{121}$Te activity
obtained from the 1-parameter fit (which takes into account all
known information) will be used to estimate the production rate.

\begin{table}
\centering
\begin{tabular}{lcccc}
\hline
 Fit & $A_{0,m}$    &  $A_{0,g}$  & $T_{1/2,m}$  &  $T_{1/2,g}$ \\
 & (kg$^{-1}$d$^{-1}$) & (kg$^{-1}$d$^{-1}$)  & (d) & (d)\\ \hline
4 parameters   & 76$\pm$15  &   159$\pm$33 &  122$\pm$27 &  10.5$\pm$5.4\\
2 parameters &59.3$\pm$3.9& 134$\pm$13 &  - & -\\
1 parameter &-  &  110$\pm$12 &  - & -\\
\hline
\end{tabular}
\caption{\label{resultste121} Initial activities underground and
half-lives derived for the metastable and ground states of
$^{121}$Te by fitting the signal plotted in figure~\ref{fitte121} to
the dependence given by eq.~(\ref{ate121}) with different sets of
free parameters (see text).}
\end{table}

\subsubsection{$^{22}$Na}

For $^{22}$Na, having a much longer mean life than I and Te
isotopes, the evolution in time of the identifying signal has not
been studied; signal integrated along the available data in set III
(88.12~days) has been considered instead.

\begin{figure}
 \centering
 \includegraphics[width=10cm]{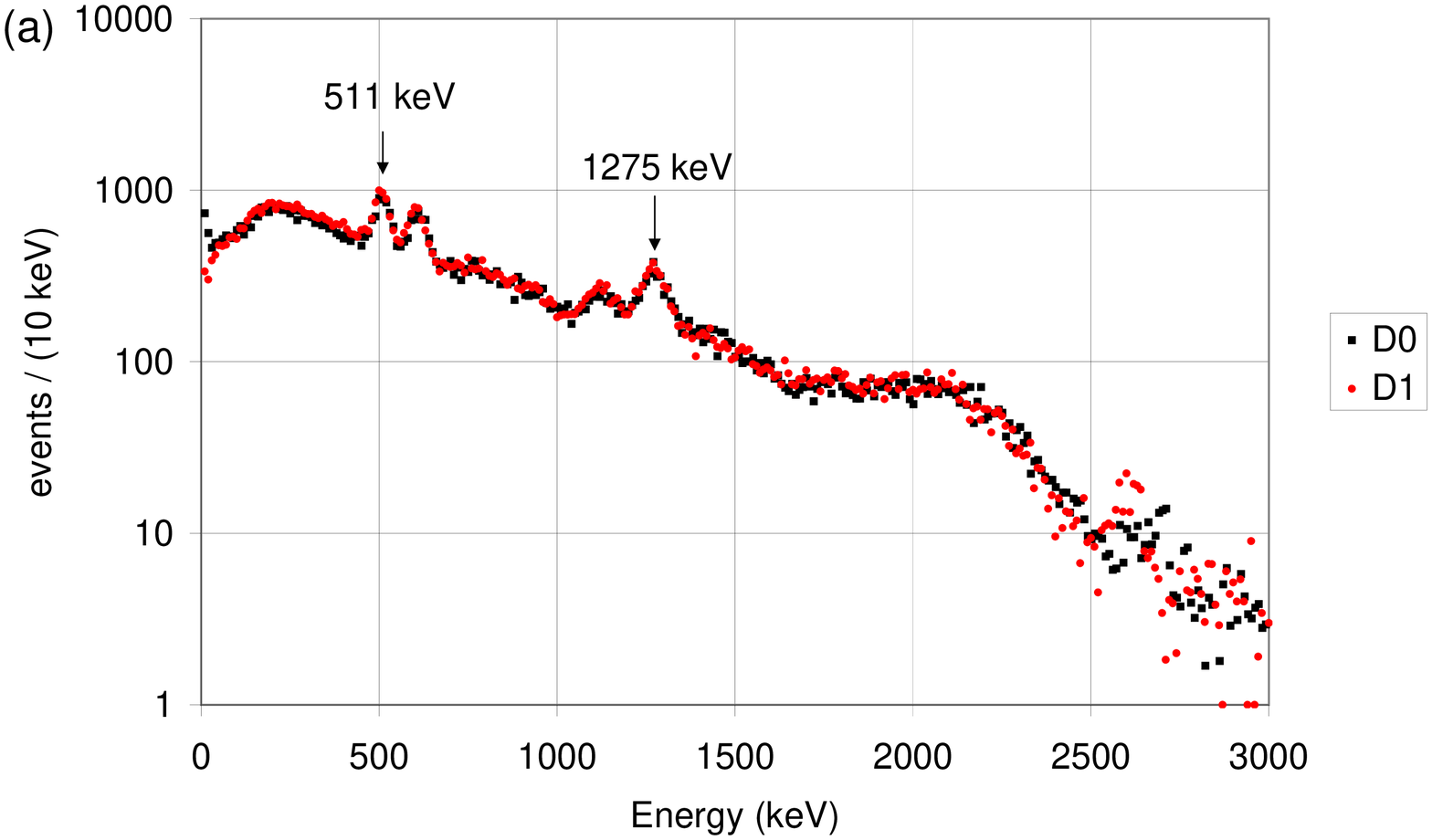}
\includegraphics[width=7cm]{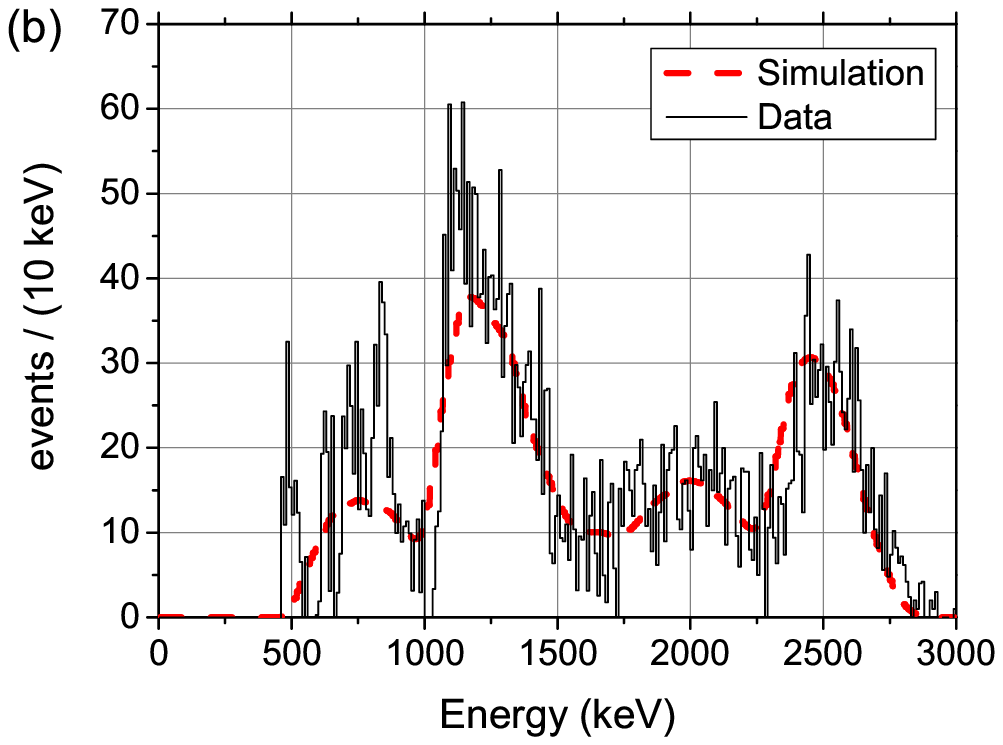}
\includegraphics[width=7cm]{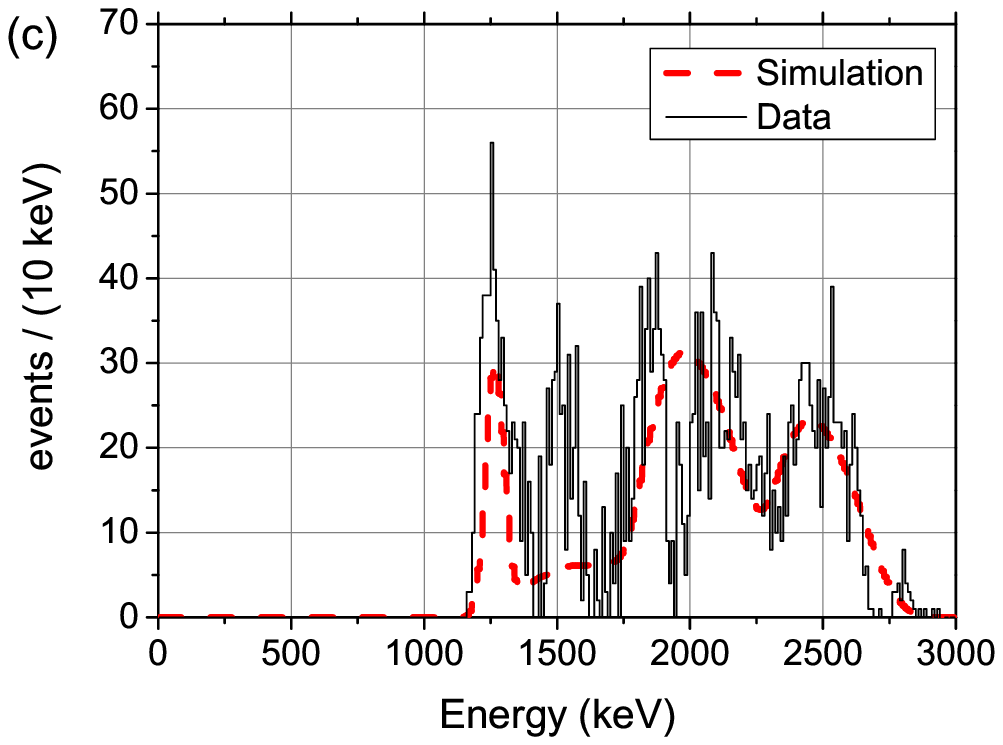}
 \caption{(a): Energy spectra of D0 (black squares) and D1 (red circles) registered for coincidences between them for data set III.
(b) and (c): Spectrum of the summed energy of D0 and D1 for events
in a $\pm$1.4$\sigma$ window around the 511 (b) and 1275~keV (c)
peaks in any of the coincidence spectra. $^{22}$Na signature has
been evaluated from the region between 2300 and 2900~keV
(corresponding to full absorption of its positron and gamma
emissions). Measured spectra are compared with simulations
normalized with the deduced $^{22}$Na activity.}
  \label{spcna22}
\end{figure}

Figure~\ref{spcna22} (a) shows the energy spectra of D0 and D1
registered for coincidences between them for data set III. Two
complementary signatures have been considered: events selected by a
511~keV (1275~keV) energy deposition in one detector (fixing a
window at $\pm$1.4$\sigma$) have been used to build total deposited
energy spectra and then, only the region between 2300 and 2900~keV
(corresponding to full absorption of its positron and gamma
emissions) has been used for the rate estimate. Figure~\ref{spcna22}
(b,c) presents these total energy spectra (summing D0 and D1 energy)
obtained for 511 or 1275~keV peak. The effect of coincidences not
due to $^{22}$Na has been taken into account subtracting their
contribution in the 2300-2900 keV region, evaluated from events in a
window on the left of the 511~keV peak and on the right of the
1274~keV peak. From the net number of events registered in the
2300-2900 keV region and the estimated $I\epsilon$ value reported at
table~\ref{isotopes} the activity of $^{22}$Na at measuring time has
been evaluated and then the initial activity underground $A_{0}$
deduced, using its half-life from table~\ref{isotopes}.
Table~\ref{na22} presents the results, for both 511 and 1275~keV
windows; since they are compatible, they have been properly
averaged. In figure~\ref{spcna22} (b,c) the measured spectra of the
summed energy of D0 and D1 are compared with the corresponding
simulations normalized with the deduced $^{22}$Na activity; the
spectra shapes follow reasonably the expectation, somehow validating
the method applied to single out $^{22}$Na contribution.

\begin{table}
\centering

\begin{tabular}{lcc}
\hline Signal (keV) & A (kg$^{-1}$d$^{-1}$) from data set III & $A_{0}$ (kg$^{-1}$d$^{-1}$) \\ 
\hline 511  & 109.0$\pm$4.3& 158.2$\pm$6.3 \\
1275   & 111.8$\pm$5.5 & 162.2$\pm$8.0 \\
mean  & 110.0$\pm$3.4 &159.7$\pm$4.9 \\ \hline
\end{tabular}
 \caption{\label{na22} Measured activity of $^{22}$Na from data set III and the deduced initial activity
underground $A_{0}$, considering the identifying signal for
coincident events leaving 511 or 1275~keV at one detector (see
text). These activities independently estimated have been properly
averaged to derive the final result.}
\end{table}

\subsection{Estimate of production rates at sea level}
\label{Rp}

ANAIS-25 crystals were built in Alpha Spectra facilities in Grand
Junction, Colorado. The history of the starting material used to
produce the NaI powder from which crystals are grown is unknown;
this exposure along geological time scales must be relevant only for
very long-lived isotopes like $^{129}$I or $^{123}$Te, for those
having shorter lifetimes saturation would have been probably reached
independently from previous exposure history while purifying,
growing the crystals and building the detectors (except maybe for
$^{22}$Na). Colorado is placed at a quite high altitude;
consequently, a very important correction factor $f$ to the cosmic
neutron flux at New York City coordinates is reported at
\cite{ziegler} for Colorado locations due to altitude and
geomagnetic rigidity (4.11 and even 12.86 for Denver and Leadville,
respectively). Afterwards, detectors were transported to LSC by boat
and by road, being the travel duration about one month, most of the
time at sea level. Therefore, the initial activities underground
derived in section~\ref{A0} cannot be directly considered the
saturation activities (i.e. production rates, $R_{p}$) at sea level.
However, in the following a method to estimate the production rates
in our crystals, based on a few reasonable assumptions, is proposed.

Considering that saturation activity is reached at a given place
characterized by a correction factor $f$ to the cosmic neutron flux
at sea level, the number of nuclei cosmogenically induced for a
particular isotope is $N_{sat}=f R_{p}/\lambda$. If the material is
then exposed to cosmic rays at sea level for a time $t$, it is
straightforward to derive the posterior evolution of the
corresponding activity by solving the usual differential equation
with simultaneous production and decay of a nuclear species for an
initial number of nuclei $N_{sat}$:
\begin{equation}
A(t) = R_{p} [1 + (f-1) \exp(-\lambda  t)] \label{eqrp}
\end{equation}

The production rate $R_{p}$ of an isotope can be obtained from
eq.~(\ref{eqrp}) using its initial activity underground, obtained in
section~\ref{A0}, and fixing the correction factor $f$ and exposure
time at seal level $t$. Since the geographic coordinates of Grand
Junction are similar to those of Denver and Leadville, their
correction factors $f$ have been adjusted to Grand Junction altitude
(4593~ft$=$1397~m) following the method proposed in \cite{ziegler}
(see table~\ref{factorf}); the average of both estimates has been
taken hereafter as $f$ value. Table~\ref{resultsrp} summarizes the
production rates obtained in this way for $f=$3.6$\pm$0.1,
$t=$30$\pm$5~days, $\lambda$ from T$_{1/2}$ values in
table~\ref{isotopes} and the $A_{0}$ values presented in
tables~\ref{finalfits} (considering the isotope decay along the data
taking) and \ref{na22}. Reported uncertainties include contributions
from those derived for $A_{0}$ and also from $\lambda$, $f$ and $t$.
The relative uncertainty quantified in the production rate is
$\leq$5\% (except for $^{126}$I); assuming much larger uncertainties
in $f$ and $t$, as for instance $f=$3.6$\pm$0.5 and
$t=$30$\pm$10~days, increases the relative uncertainties in the
production rates to $\sim$15\%. Contribution from uncertainty in
half-lives is negligible.

\begin{table}
\centering
\begin{tabular}{lcc}
\hline & Denver & Leadville \\ \hline
$f$ \cite{ziegler}& 4.11 & 12.86 \\
altitude in ft (m) & 5130-5690 (1564-1731) & 10152 (3094) \\
flux ratio to Grand Junction & 1.18 & 3.47\\
$f$ at Grand Junction & 3.5 & 3.7\\ \hline
\end{tabular}
 \caption{\label{factorf} Estimate of the correction factor $f$ for the
cosmic neutron flux at Grand Junction from the factors at Denver and
Leadville. The ratio between the flux at these locations and Grand
Junction due to different altitude has been derived following
\cite{ziegler}; the altitude of Grand Junction is 4593~ft (1397~m).}
\end{table}

\begin{table}
\centering
\begin{tabular}{lc}
\hline Isotope & $R_{p}$ (kg$^{-1}$d$^{-1}$) \\ 
\hline $^{126}$I  &      283$\pm$36 \\ 

$^{125}$I &    220$\pm$10 \\

$^{127m}$Te &    10.2$\pm$0.4 \\ 

$^{125m}$Te &    28.2$\pm$1.3 \\

$^{123m}$Te &   31.6$\pm$1.1 \\ 

$^{121m}$Te &    23.5$\pm$0.8 \\

 $^{121}$Te  &  9.9$\pm$3.7 \\ 
 $^{22}$Na &  45.1$\pm$1.9  \\ \hline
\end{tabular}
 \caption{\label{resultsrp} Production rates at sea level $R_{p}$ deduced
for all the identified cosmogenic isotopes (in kg$^{-1}$d$^{-1}$).
Values for I and metastable Te isotopes have been obtained from
eq.~(\ref{eqrp}) using the initial activities underground $A_{0}$
presented in table~\ref{finalfits}, calculated as the average values
obtained for different time intervals. Rate for ground state of
$^{121}$Te has been deduced from eq.~(\ref{ate121sea}) using the
activity in table~\ref{resultste121} from the 1-parameter fit. For
$^{22}$Na, eq.~(\ref{eqrp}) and the mean result in table~\ref{na22}
have been considered.}
\end{table}

For $^{121}$Te, assuming an exposure to cosmic rays in Colorado
until reaching saturation followed by a short exposure at sea level
(as made for the other cosmogenic products), the estimate of the
production rate from the measured initial activity underground is
not so easy due to the combination of direct production and
production through the metastable state. In the following, $R_{p,g}$
and $R_{p,m}$ correspond to the production rates at sea level of the
ground and metastable states of $^{121}$Te.

At an altitude with a correction factor $f$ for the cosmic neutron
flux, the number of nuclei $N$ of each species is given by the
system of coupled equations:
\begin{equation}
\label{nte121}
\begin{array}{rcl}
 dN_{m} &=& f R_{p,m} dt - \lambda_{m} N_{m} dt \\
 dN_{g} &=& f R_{p,g} dt - \lambda_{g} N_{g} dt + P_{IT} \lambda_{m} N_{m}
 dt
 \end{array}
\end{equation}
The system in eq.~(\ref{nte121}) can be solved assuming that
initially $N_{g}=N_{m}=0$. Then, saturation values (for
$t\rightarrow\infty$) are:
\begin{equation}
\label{nsatte121}
\begin{array}{rcl}
N_{m,sat} &=& \frac{f R_{p,m}}{\lambda_{m}}\\
N_{g,sat} &=& \frac{f (R_{p,g} + P_{IT} R_{p,m})}{\lambda_{g}}
 \end{array}
\end{equation}

At sea level, the system of coupled equations giving the number of
nuclei of ground and metastable isotopes is like that in
eq.~(\ref{nte121}) with $f=$1
and can be solved assuming initially the saturation values from
eq.~(\ref{nsatte121}). Then, the activity for the metastable state
is given by eq.~(\ref{eqrp}), whereas activity for the ground state
is obtained as:
\begin{equation}
A_{g}(t) = (R_{p,g} + R_{p,m} P_{IT}) [1 + (f-1)
\exp(-\lambda_{g}t)] + R_{p,m} P_{IT} (f-1)
[\exp(-\lambda_{m}t)-\exp(-\lambda_{g}t)]
\frac{\lambda_{g}}{\lambda_{g}-\lambda_{m}} \label{ate121sea}
\end{equation}


Table~\ref{resultsrp} presents the production rate at sea level
$R_{p,g}$, derived from eq.~(\ref{ate121sea}), using the
corresponding initial activity underground deduced in
table~\ref{resultste121} from the 1-parameter fit, the values of
$\lambda_{m}$, $\lambda_{g}$ and $P_{IT}$ and the values of $f$ and
$t$ already assumed for the other cosmogenic products. Uncertainty
in the activity give the main contribution to the uncertainty in the
production rate.


\section{Calculation of production rates at sea level}
\label{cal}

As pointed out in section~\ref{intro}, cosmogenic activation of
materials is an hazard for experiments looking for rare events and
therefore it must be properly assessed in the background models of
this type of experiments. Production rates are the main input to
compute the induced activities of the relevant isotopes. Since
direct estimates of the production rates, as those presented here
for NaI, are not available for all materials and products, in many
cases production rates must be evaluated using eq.~(\ref{eqrate})
from the cosmic ray flux $\phi$ and the production cross sections
$\sigma$, both dependent on the energy E of cosmic particles:
\begin{equation}
R_{p} \propto\int\sigma(E)\phi(E)dE \label{eqrate}
\end{equation}
This calculation method is strongly dependent on the excitation
functions and assumed cosmic ray spectrum. In order to validate
different approaches to such a calculation, production rates at sea
level for all the cosmogenic products identified and quantified in
ANAIS-25 crystals will be evaluated and then compared with the
values obtained from the measurements presented in
section~\ref{ana}. Although not observed in ANAIS-25 data mainly due
to its short half-life ($T_{1/2}=$4.1760$\pm$0.0003~days), $^{124}$I
has been considered in this study too. Production rate of
$^{123}$Te, having low energy emissions dangerous for dark matter
searches, has been also evaluated. As proposed in \cite{cebrian},
first, all the available information on the relevant production
cross sections by neutrons and protons from both measurements and
calculations will be presented; then, the best description of the
excitation functions of products over the whole energy range will be
chosen (section~\ref{excf}). Finally, production rates considering a
particular cosmic ray spectrum (section~\ref{crs}) and the selected
excitation functions will be calculated and compared with the
experimental values (section~\ref{prodr}).

\subsection{Excitation functions}
\label{excf}

The excitation function for the production of a certain isotope by
nucleons in a target over a wide range of energies (from some MeV up
to several GeV) can be hardly obtained only from experimental data,
since the measurements of production cross-sections with beams are
demanding. The use of computational codes is therefore mandatory,
although experimental data are essential to check the reliability of
calculations. The difficulties for properly describing excitation
functions are discussed for instance at \cite{reedy}.

Most of the considered cosmogenic isotopes are induced by
interaction on $^{127}$I; only $^{22}$Na is produced on $^{23}$Na.
Both $^{127}$I and $^{23}$Na have a 100\% isotopic abundance in
natural iodine and sodium respectively.
Figures~\ref{sigmai126}-\ref{sigmana22} in the appendix show the
information compiled for the excitation functions of the cosmogenic
products induced on $^{127}$I or $^{23}$Na, taken from different
sources:
\begin{itemize}
\item The EXFOR database (CSISRS in US) \cite{exfor} provided most
of the experimental data. Other ones have been obtained from
\cite{dyer}, \cite{prons} and \cite{vrza}.
\item Libraries MENDL-2 and MENDL-2P (Medium Energy Nuclear Data Library)
\cite{mendl} gave results obtained by Monte Carlo simulation of the
hadronic interaction between nucleons and nuclei using codes of the
ALICE family, for neutrons up to 100~MeV and protons up to 200~MeV.
\item At library TENDL-2013 (TALYS-based Evaluated Nuclear Data Library)
\cite{tendl}, cross sections obtained with the TALYS nuclear model
code system were found for neutrons and protons up to 200~MeV. An
outstanding feature of this library is that production of metastable
and ground states is evaluated separately.
\item The YIELDX routine was used to compute production cross section at the
whole energy range. This routine implements the most updated version
of the popular Silberberg \& Tsao semiempirical formulae
\cite{tsao1}-\cite{tsao3} giving nucleon-nucleus cross sections for
different reactions. They can be used for a wide range of targets
and products at energies above 100~MeV; since they are based only on
proton-induced reactions, cross section must be assumed to be the
same for neutrons and protons.
\item Library HEAD-2009 (High Energy Activation Data) \cite{head2009}
merges data for neutrons and protons up to 1~GeV. Here, data from
HEPAD-2008 (High-Energy Proton Activation Data) sub-library have
been considered, obtained using a selection of models and codes
(CEM03.01, CASCADE/INPE, MCNPX~2.6.0,\dots) dictated by an extensive
comparison with EXFOR data.

\end{itemize}
At low energies (LE, below 200~MeV) excitation functions for protons
and neutrons are clearly different, from both MENDL and TENDL-2013
data; therefore, it is very important to quantify cosmogenic yields
considering cross sections specifically evaluated for neutrons. For
I isotopes, experimental data on cross sections are abundant and are
in general in reasonable agreement with the excitation functions
provided by MENDL, TENDL-2013, YIELDX and HEAD-2009. However, for Te
isotopes, experimental information is scarce. It is worth noting
that some measurements distinguish production cross sections for
ground (G) or metastable (M) states of Te isotopes, but this is not
the case for the considered calculations except for TENDL-2013.

To quantify the agreement between the available experimental data
and the different calculations helps to choose the best description
of the excitation functions. For that, deviation factors $F$ have
been evaluated following a definition typically used in the
literature:
\begin{equation}
F=10^{\sqrt{d}},\hspace{0.5cm}
d=\frac{1}{N}\sum_{i}(\log\sigma_{exp,i}-\log\sigma_{cal,i})^{2}
\label{eqF}
\end{equation}
\noindent being $N$ the number of pairs of experimental and
calculated cross sections $\sigma_{exp,i}$ and $\sigma_{cal,i}$ at
the same energies. Parameter $d$ is the mean square logarithmic
deviation. Table~\ref{df} presents the obtained values for each
isotope and considering all of them altogether; only proton data can
be considered for HEAD-2009 and YIELDX while data for neutrons are
used for MENDL-2 and TENDL-2013 calculations. Only an abnormally
large deviation factor above 10 has been found for $^{127m}$Te and
HEAD-2009 calculation, predicting very low production cross sections
well below available measurements for both metastable and ground
states. Calculations at high energy show slightly better deviation
factors than at low enegies; in the overall results not very
significant differences are found between HEAD-2009 and YIELDX nor
between MENLD-2 and TENDL-2013.

\begin{table}
\centering
\begin{tabular}{l|cc|cc|cc|cc}
\hline

Isotope &HEAD-2009 &  & YIELDX  &  &  MENDL-2 &  & TENDL-2013(n)&   \\
  &  F  & N &  F  & N & F & N & F & N \\\hline
$^{126}$I &   1.3 &18 & 1.2& 25 & 1.4 &37 & 1.4 &38\\

$^{125}$I  &  1.3 &8  & 2.1& 12& 1.7& 1 & 1.1& 1\\

$^{124}$I   & 1.4 &12 &1.5 &19 & 2.9 &6& 3.9 &6\\

$^{127m}$Te & 11.5& 2& & & 1.9&3 & 1.5 & 3 \\

$^{125m}$Te & & & & & & & 5.6 & 1 \\

$^{123m}$Te &1.2 &4 &  2.4 &6&  & & & \\

$^{121m}$Te & 1.1& 4 &1.8& 6& & \\

$^{121}$Te  & 1.7 &4 &  1.4 &6 & & \\ 

$^{22}$Na & & & 1.2 & 19& & & 1.8 & 107 \\ \hline
 all  &   1.4* &50  &1.5 &93&  1.6 &47 & 1.8 & 160 \\ \hline
\end{tabular}
(*)Excluding the large deviation factor obtained for $^{127m}$Te.
\caption{\label{df} Deviation factors $F$ calculated following
eq.~(\ref{eqF}) between measured production cross-sections and
results from HEAD-2009 (150-1000~MeV), YIELDX ($>$100~MeV), MENDL-2
($<$100~MeV) and TENDL-2013 for neutrons ($<$200~MeV). The number
$N$ of pairs used in the comparison is indicated for each case.
Overall values of $F$ have been obtained considering data for all
the isotopes.}
\end{table}

Different descriptions of the excitation functions have been
analyzed for the calculation of production rates, based on the
available data and calculations at the low and high energy ranges:
\begin{itemize}
\item For iodine isotopes, thanks to the large amount of
experimental data available, the description could be fixed for each
isotope following the lowest deviation factors at low and high
energies according to table~\ref{df}.
\item For tellurium isotopes, TENDL-2013 data for neutrons were always considered below 200~MeV, since
they have been calculated separately for metastable states;
HEAD-2009 data, having better deviation factors than YIELDX, were
used above that energy for most of the isotopes. For $^{127m}$Te, it
was preferred to extrapolate the TENDL-2013 value at 200~MeV to
higher energies. The branching ratios for the production at ground
or metastable state given by TENDL-2013 data at 200~MeV ($\sim$0.7
for metastable production for all the considered isotopes) have been
applied to data at higher energies, which do not distinguish between
ground or metastable final states.
\item Regarding $^{22}$Na, the only description possible was using TENDL-2013 at low energies and YIELDX results at high energies.
\end{itemize}

\subsection{Cosmic rays spectrum}
\label{crs}
\begin{figure}
 \centering
 \includegraphics[width=10cm]{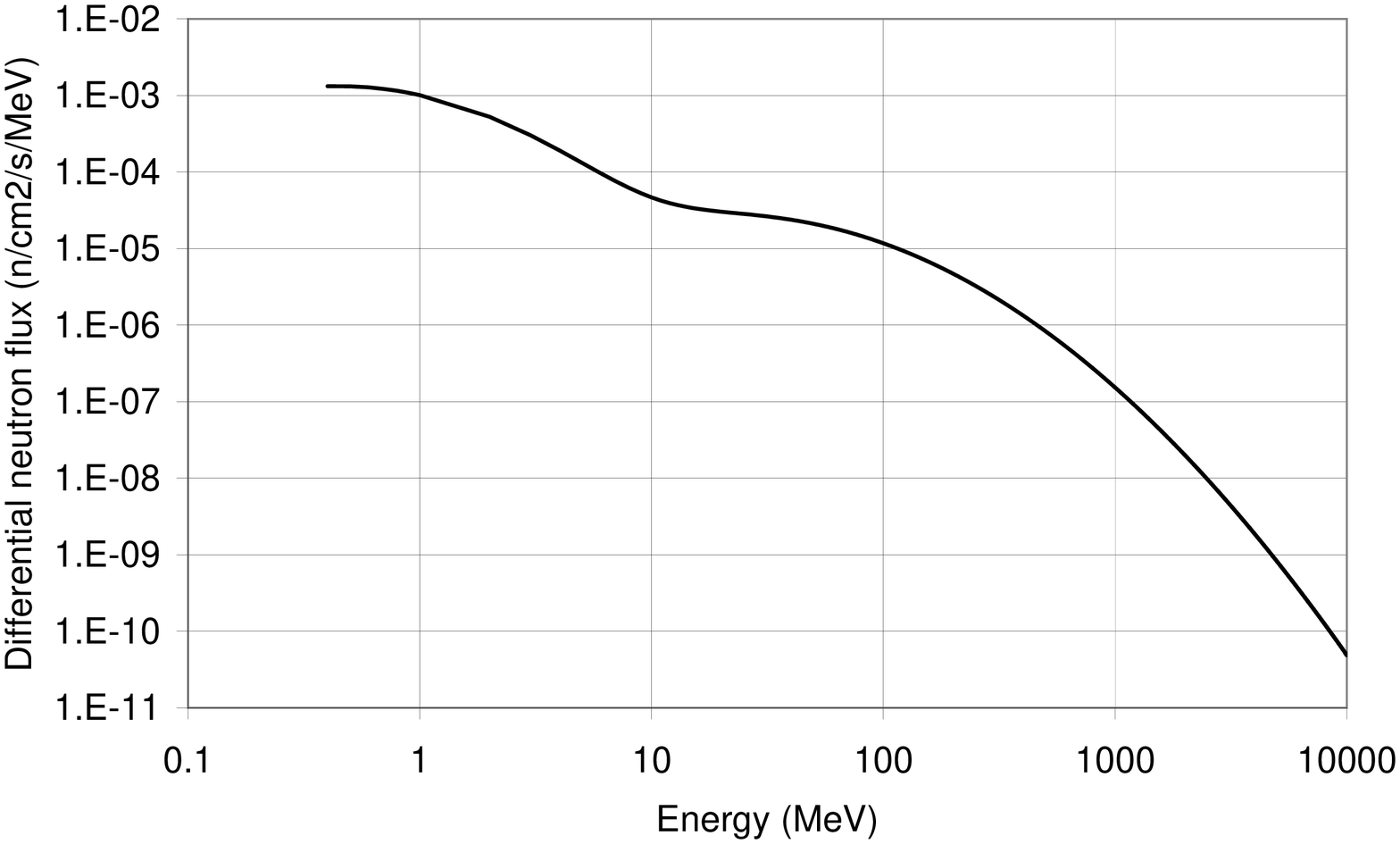}
 \caption{Energy spectrum of cosmic neutrons following the analytic expression in \cite{gordon} (for conditions of New York City at sea level) which has been considered in this work to calculate production rates of cosmogenic isotopes.}
  \label{nspectrum}
\end{figure}

Following eq.~(\ref{eqrate}), the other ingredient required to
evaluate production rates, together with the excitation function, is
the cosmic ray spectrum. Different forms of the neutron spectrum at
sea level have been used in cosmogenic activation studies, like the
historical Hess spectrum \cite{hess}, that proposed by Lal \& Peters
\cite{lal} or the parametrization based on a compilation of
measurements presented by Ziegler \cite{ziegler}. In \cite{gordon},
a set of measurements of cosmic neutrons on the ground across the US
was accomplished using Bonner sphere spectrometers; an analytic
expression fitting data for energies above 0.4~MeV was proposed and
has been considered in this work. Figure~\ref{nspectrum} shows the
energy spectrum given by this description of cosmic neutrons;
applied to conditions of New York City at sea level, the integral
flux from 10~MeV to 10~GeV is 3.6$\times 10^{-3}$cm$^{-2}$s$^{-1}$.

\subsection{Production rates}
\label{prodr}
Table~\ref{calrp} presents the obtained production rates for the
analyzed cosmogenic isotopes considering the selected excitation
functions. Results are given per day and per kg of NaI.
Contributions from the LE and HE descriptions of the cross sections
are shown together with the total rate. For most of the isotopes,
the dominant contribution comes from the low energy region; for
instance, for iodine isotopes it accounts for $\sim$90\% of the
rate.

\begin{table}
\centering
\begin{tabular}{lccccc}
\hline Isotope & Excitation function &  Calculated rate & & Experimental rate & Cal$/$Exp \\
& LE+HE & LE+HE & total & & \\ \hline

$^{126}$I & MENDL-2+YIELDX   & 250.0+47.0 &297.0 & 283$\pm$36  & 1.1 \\

$^{125}$I & TENDL-2013+HEAD-2009 & 230.2+12.1 & 242.3  & 220$\pm$10 & 1.1  \\

$^{124}$I & MENDL-2+HEAD-2009 & 113.6+22.3 & 135.9  & & \\

$^{127m}$Te &TENDL-2013+extrapolation   &  6.9+0.2 &7.1 &  10.2$\pm$0.4 & 0.7 \\

$^{125m}$Te &TENDL-2013+HEAD-2009   &  38.4+3.5 &41.9 &  28.2$\pm$1.3 & 1.5 \\

$^{123m}$Te & TENDL-2013+HEAD-2009 & 29.5+3.7 &33.2  &  31.6$\pm$1.1 & 1.1 \\

$^{123}$Te & TENDL-2013+HEAD-2009 &  8.8+1.4 &10.2 & & \\

$^{121m}$Te & TENDL-2013+HEAD-2009 & 19.1+4.7  &23.8  &  23.5$\pm$0.8 &1.0 \\

$^{121}$Te & TENDL-2013+YIELDX  & 5.8+2.6 & 8.4  & 9.9$\pm$3.7 & 0.8 \\

$^{22}$Na & TENDL-2013+YIELDX &  43.2+10.4 &53.6 & 45.1$\pm$1.9 &  1.2\\
\hline

\end{tabular}
 \caption{\label{calrp} Production rates (in kg$^{-1}$d$^{-1}$)
calculated from eq.~(\ref{eqrate}) using different selections of the
excitation function (indicated in the second column) showing the
contributions from LE and HE ranges and the total rate. Production
rates obtained experimentally in table~\ref{resultsrp} are shown for
comparison. Last column presents the ratio between the calculated
and the experimental rates.}
\end{table}

Table~\ref{calrp} shows also the production rates deduced
experimentally in section~\ref{Rp} for comparison. Ratios between
the calculated and experimental rates are given too. Most of the
calculated production rates are larger than the experimental ones.
Good agreement has been found for I isotopes, having cross sections
well validated not only at high energy with protons but also at low
energies with neutrons. The largest deviation is obtained for
$^{125m}$Te; the lack of experimental data in the whole energy range
(see Fig.~\ref{sigmate125}) makes difficult try to improve the
selection of the excitation function. For $^{22}$Na, there is an
important step in the description of the excitation curve at LE and
HE regions (see Fig.~\ref{sigmana22}) and then contribution from HE
could be overestimated and/or that from LE undervalued, justifying
the difference between the calculated and experimental rates; in any
case, they are compatible with the calculated maximum rate of
$\simeq$ 100 kg$^{-1}$d$^{-1}$ at sea level quoted in
\cite{damanima2008}. The fact of observing a rate below the
calculation could indicate that saturation was indeed not reached
for this isotope. There is a similar mismatch in the description of
the excitation function at LE and HE regions for $^{121}$Te (see
Fig.~\ref{sigmate121}); but in this case re-scaling the curves would
give production rates still compatible with the observed value.

Given the discrepancies found between calculated and experimental
production rates, it can be concluded that calculations of
activation yields in sodium iodide for ultra-low background
experiments are precise enough, provided the used cross sections are
well validated along all the relevant energy range.

\section{Contribution to detector background levels}
\label{contri}

The impact of the cosmogenic isotopes produced in NaI(Tl) crystals
on the background levels at the energy regions relevant for rare
event searches, mainly the direct detection of WIMPs, has been
assessed by means of Monte Carlo simulation. The same Geant4
application developed for the ANAIS-25 set-up and used to determine detection
efficiencies (see section \ref{ana}) has been used. The Radioactive
Decay Model in Geant4 has been considered to reproduce the
radioactive decays of the products.

\begin{figure}
 \centering
 \includegraphics[width=8cm]{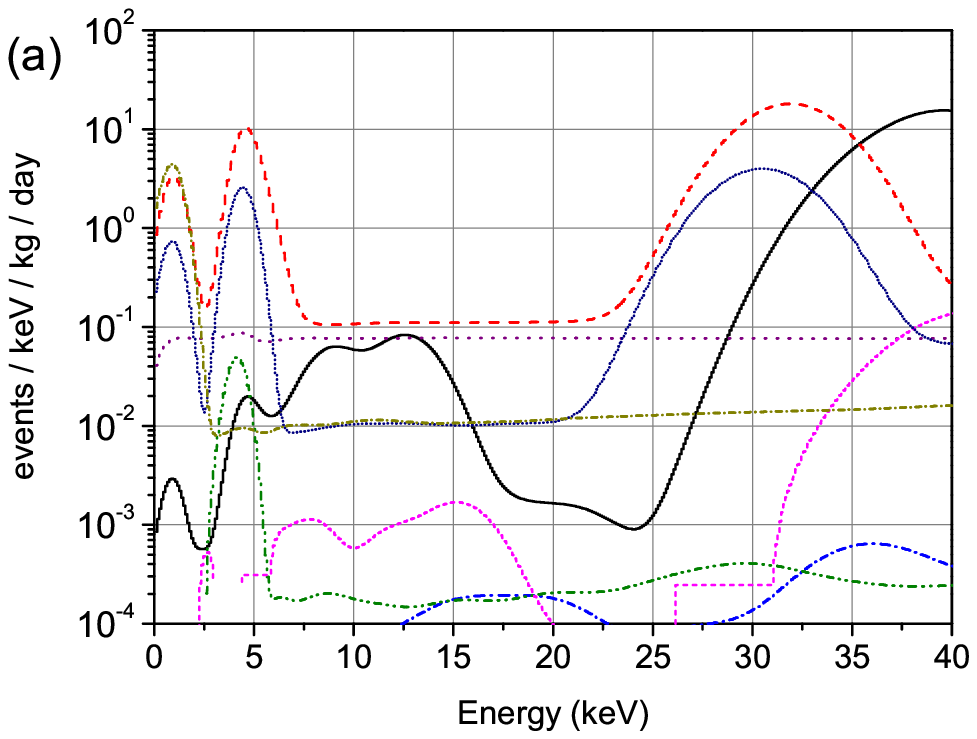}
 \includegraphics[width=8cm]{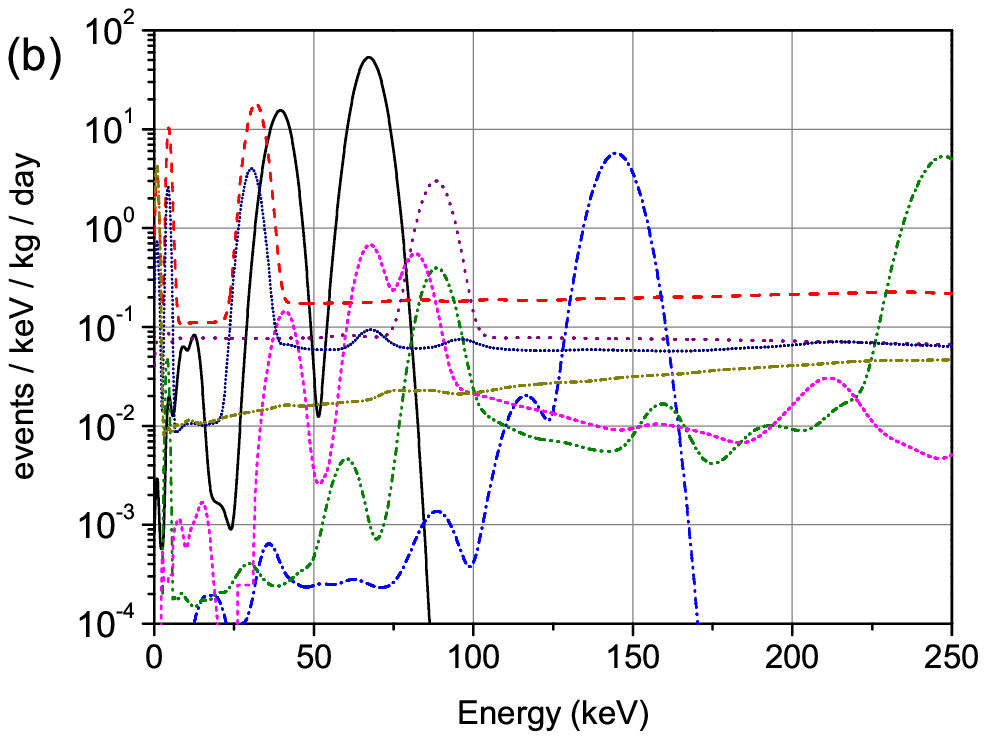}
 \includegraphics[width=8cm]{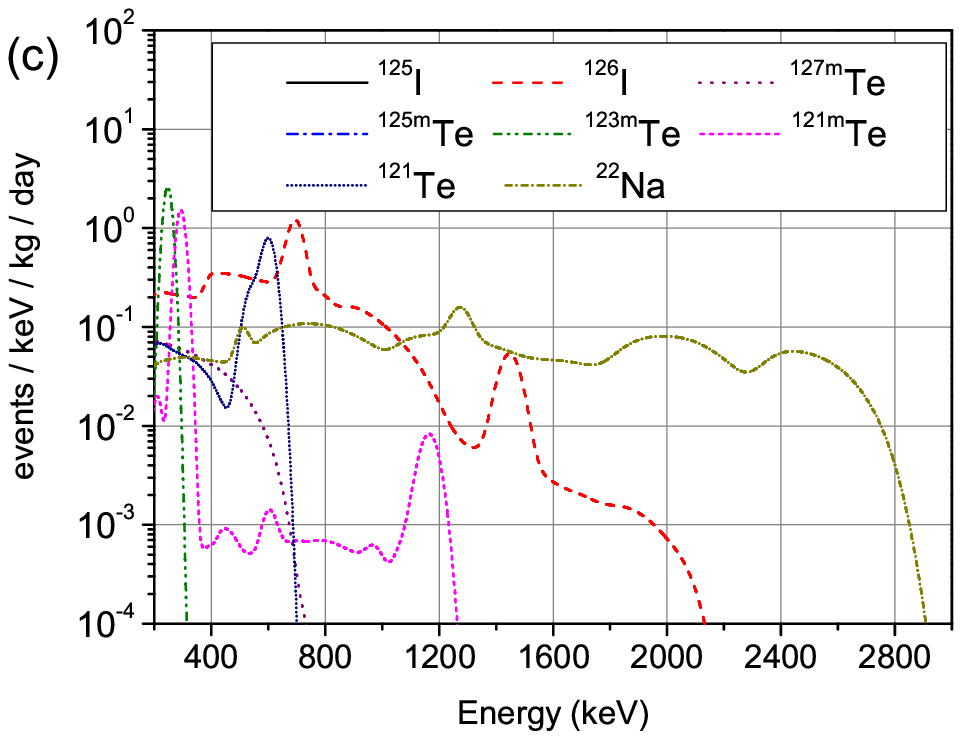}
 \caption{Background contribution from each identified cosmogenic isotope obtained by Geant4 simulation of ANAIS-25 set-up assuming the measured initial activities $A_{0}$ in three different energy ranges: (a) very low energy (below 40 keV), (b) medium energy (below 250 keV) and (c) high energy (from 200 to 3000 keV). The same legend applies to the three plots.}
  \label{simdesglose}
\end{figure}

\begin{figure}
 \centering
 \includegraphics[width=8cm]{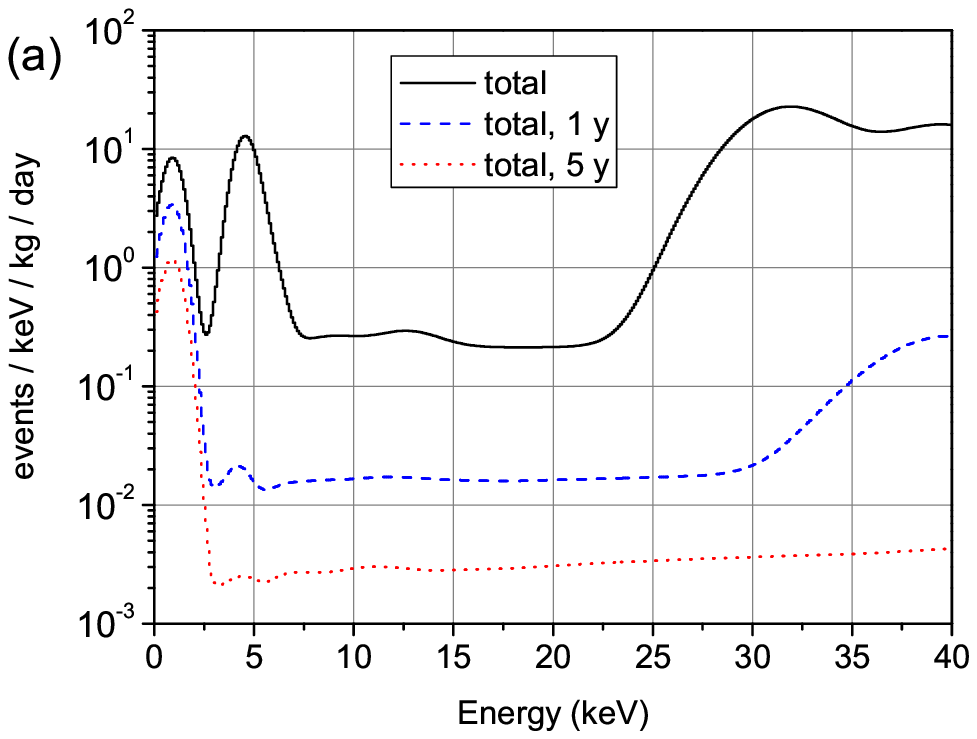}
 \includegraphics[width=8cm]{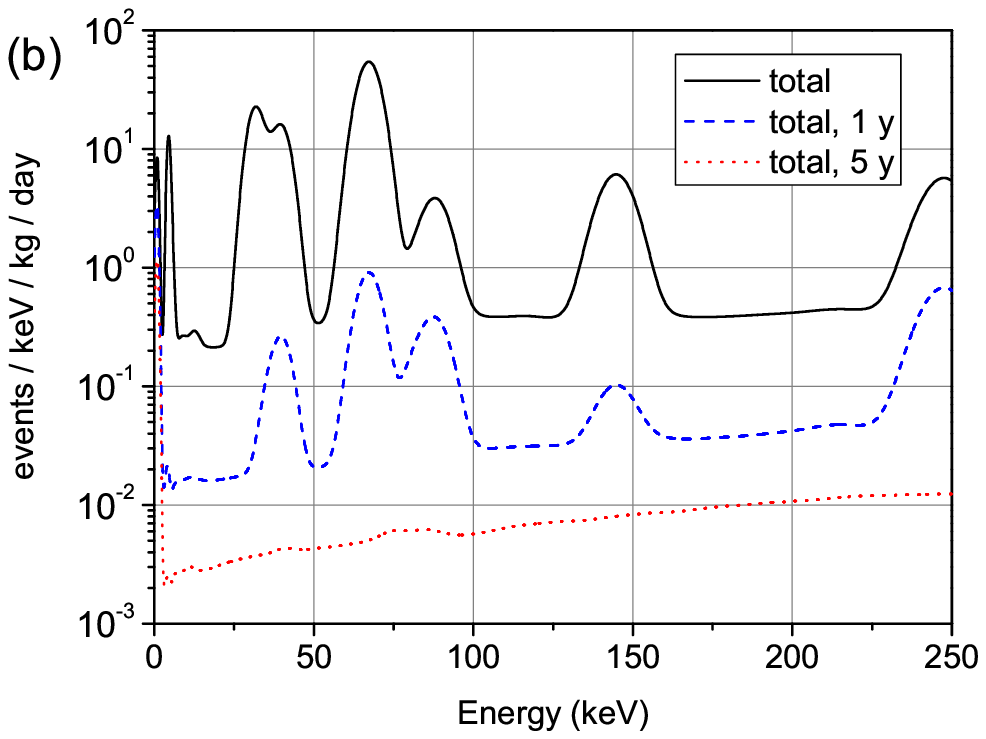}
 \includegraphics[width=8cm]{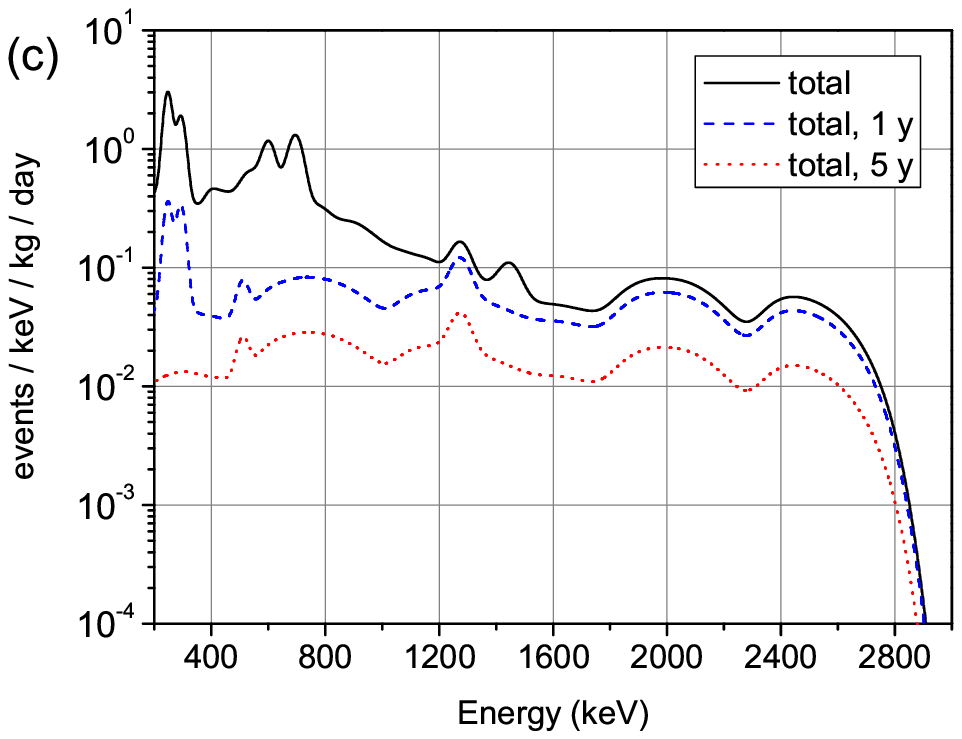}
 \caption{Sum of the background contributions depicted in Figure \ref{simdesglose} (black). The evolution of this cosmogenic background is shown after one  year (blue) and five years (red). $^{22}$Na contribution is dominating the latter.}
  \label{simtotal}
\end{figure}

Figure \ref{simdesglose} presents the simulated spectra for each identified
nuclide considering the measured initial activities underground
$A_{0}$ from tables \ref{finalfits}, \ref{resultste121} and \ref{na22}; three different energy
ranges are shown, including the very low energy region where dark
matter signals are expected to appear. The total contribution
from these cosmogenic isotopes has been evaluated and it is depicted
in figure \ref{simtotal}. A detailed background model of ANAIS-25 is
underway (very first results can be found in \cite{anais25}); the
inclusion of the contribution of cosmogenically produced isotopes is essential to properly reproduce the data
taken in the first months, especially at medium and low energy \cite{anais25}. As it can be observed in Figures \ref{simdesglose} and \ref{simtotal}, background rates at the level of several counts/keV/kg/day are coming from the identified isotopes.

In figure \ref{simtotal} the total contribution to the background is also compared with
the one obtained considering one and five years of cooling
underground. It can be concluded that although initially the
contribution to the detector background level is very important,
after five years the effect of cosmogenic products is completely
negligible except for $^{22}$Na. This isotope contributes to the
whole energy region up to almost 3 MeV and gives, following the
electron capture decay, a peak at 0.87 keV (the binding energy of
K-shell in neon). Below 10~keV, the estimated contribution of
$^{22}$Na from the simulation is 5.6~kg$^{-1}$d$^{-1}$ considering
the initial activity $A_{0}$; after one (five) years, this rate is
reduced to 4.3 (1.5)~kg$^{-1}$d$^{-1}$. It must be noted that a
further reduction in these values of a factor 2.5 is foreseen by
anticoincidence event selection, according to simulation of a set-up
with 20 detectors like those of ANAIS-25. As a reference,
contribution from the measured level of $^{40}$K in ANAIS-25
crystals is 5.0~kg$^{-1}$d$^{-1}$ peaked at 3.2~keV.

In this study, decay of $^{123}$Te has not been taken into account
since the upper limit set to its half-life is very large.
Nevertheless, since it could affect the very low energy region, it
has been checked that considering the production rates deduced in
table \ref{calrp} for both metastable and ground states, for an
exposure lower than $10^{10}$ y, present activity should be lower
than 0.03~kg$^{-1}$d$^{-1}$.

\section{Conclusions}
\label{con}

Cosmogenic activation in NaI(Tl) crystals has been evaluated from
the measurements taken at LSC using two 12.5~kg NaI(Tl) detectors
produced by Alpha Spectra, in Colorado (US), and then shipped to
Spain. The prompt data taking starting, the low radioactive
background conditions and the very good energy resolution of the
detectors made possible a reliable quantification of production of
isotopes with half-lives larger than ten days.

Signatures from $^{125}$I, $^{127m}$Te, $^{125m}$Te, $^{123m}$Te and
$^{121m}$Te were directly identified in the energy spectra
registered by the detectors while production of $^{126}$I,
$^{121}$Te and $^{22}$Na was revealed by analyzing coincidences
between both detectors. The evolution of the identifying signature
of all these isotopes, except $^{22}$Na, for a period of 210~days,
following the expected exponential decay, allowed to confirm the
hypothesized origin of the cosmogenic signals and to evaluate the
activities at the start of the underground storage. Signal from
$^{22}$Na was obtained from spectra taken after fifteen months.
Independent analysis performed for each detector or different data
sets and following different methodology gave compatible results
within uncertainties. Tables~\ref{finalfits}-\ref{na22} summarize
the measured initial activity underground for all the observed
cosmogenic products. Production rates at sea level were estimated
from these activities considering that saturation was reached during
detector production in Colorado (assuming a cosmic neutron flux 3.6
times higher than at sea level) and that detectors were further
exposed for 30~days during the trip by boat from US to Spain. The
hypothesis of saturation is reasonable for the short-lived isotopes
analyzed. Results are presented in table~\ref{resultsrp}. The
analysis of the evolution in time of $^{121}$Te signal was not so
easy because direct production and production from the metastable
state, also cosmogenically induced with a longer life time, were
combined. Values for the metastable isotope deduced in this analysis
were compatible with those obtained directly from the $^{121m}$Te
signature.

Calculations of production rates for all the cosmogenic isotopes
observed were carried out considering the cosmic neutron spectrum
given in \cite{gordon} and selecting carefully the excitation
functions. For each product, the available information on production
cross sections by neutrons and protons was collected, considering
measurements, results compiled in different libraries and
calculations from semiempirical formulae made on purpose. Data from
MENDL-2 or TENDL-2013 for neutrons were considered below 100 or
200~MeV while at higher energies YIELDX calculations or results from
the HEAD-2009 library were used. For the analyzed isotopes, the
overall deviation factors range from 1.4 to 1.8, being slightly
better at high energies. Calculated production rates are summarized
in table~\ref{calrp}; agreement with the rates experimentally
deduced is very good for some of the products, having excitation
functions well validated against measurements. This comparison
allows to quantify the systematic uncertainty when applying the
followed methodology to the study of other cosmogenic products.
Uncertainties in this kind of calculations come mainly from the
difficulties encountered on the accurate description of cosmic ray
spectra, on the precise evaluation of inclusive production
cross-sections in all the relevant energy range and on the good
knowledge of the material history; the low and medium energy regions
below a few hundreds of MeV are the most problematic ones since
neutron data are scarce and differences between neutron and proton
cross sections may be important.

Some of the cosmogenic isotopes observed in this study generate
emissions which could affect the very low energy region where dark
matter signal is expected to appear in NaI(Tl) detectors. The
contribution of all the identified products on the background level
of ANAIS-25 has been evaluated by Geant4 simulation, considering the
measured initial activities and those after one or five years (see
figures \ref{simdesglose} and \ref{simtotal}). Only $^{22}$Na is
relevant in the long term, affecting the whole energy range and producing a peak
around 0.9~keV. The other products have mean lives low enough to
decay in reasonable times and they will not be a problem for dark
matter detection provided crystals are left to cool underground
before starting data taking.

\acknowledgments We deeply acknowledge C.~H.~Tsao for providing the
YIELDX routine and Yu.~A.~Korovin for sharing the HEAD-2009 library.
This work has been supported by the Spanish Ministerio de Economía y
Competitividad and the European Regional Development Fund
(MINECO-FEDER) (FPA2011-23749), the Consolider-Ingenio 2010
Programme under grants MULTIDARK CSD2009-00064 and CPAN
CSD2007-00042, and the Gobierno de Aragón (Group in Nuclear and
Astroparticle Physics, ARAID Foundation and C.~Cuesta predoctoral
grant). C.~Ginestra and P.~Villar are supported by the MINECO
Subprograma de Formación de Personal Investigador. We also
acknowledge LSC and GIFNA staff for their support.

\appendix
\section{Excitation functions}

\begin{figure}[b]
 \centering
 \includegraphics[width=12cm]{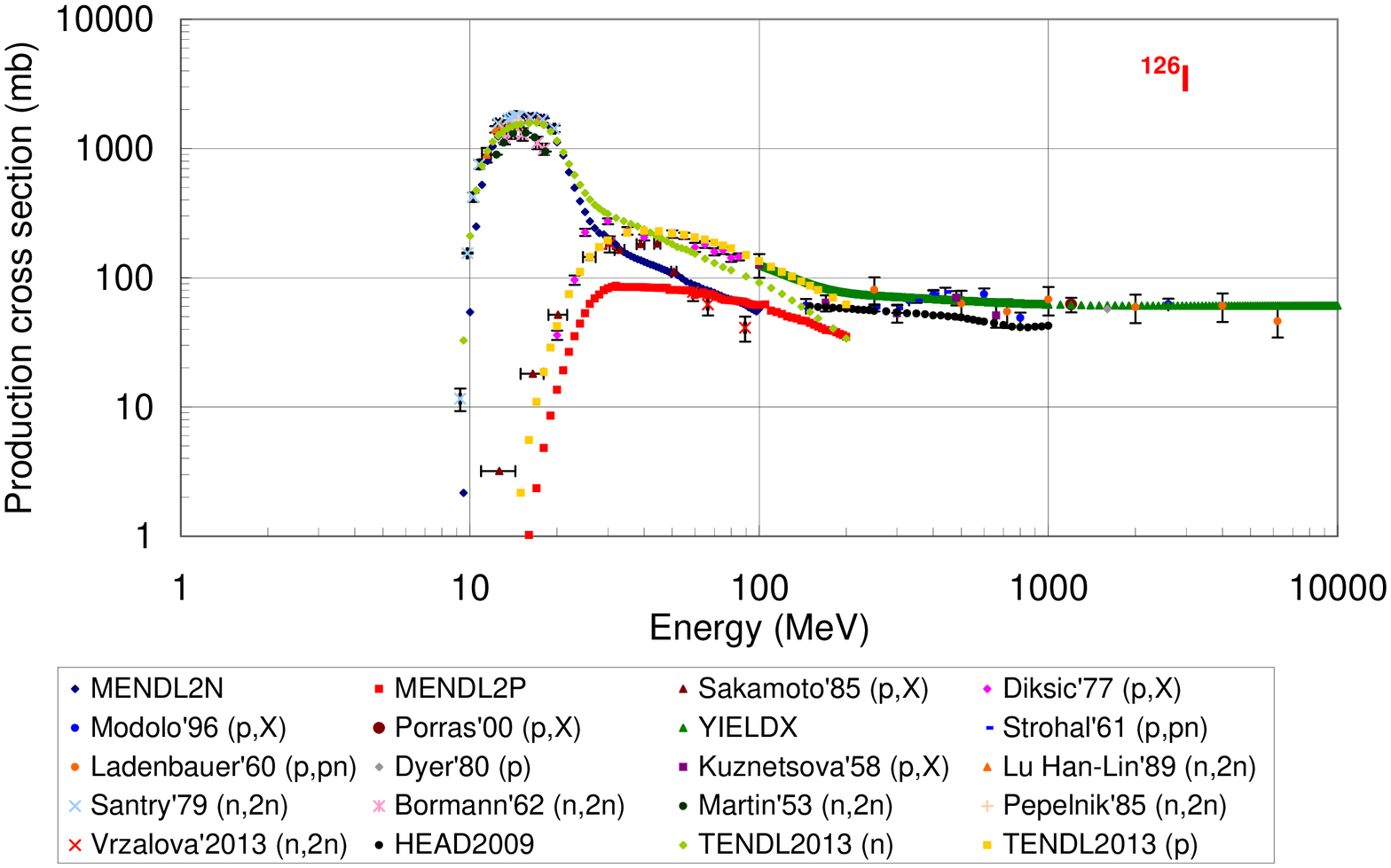} 
 \caption{Comparison of excitation functions for production of $^{126}$I on $^{127}$I by nucleons (neutrons are indicated as n and protons as p) taken from different sources (see text).}
  \label{sigmai126}
\end{figure}

\newpage

\begin{figure}
 \centering
 \includegraphics[width=12cm]{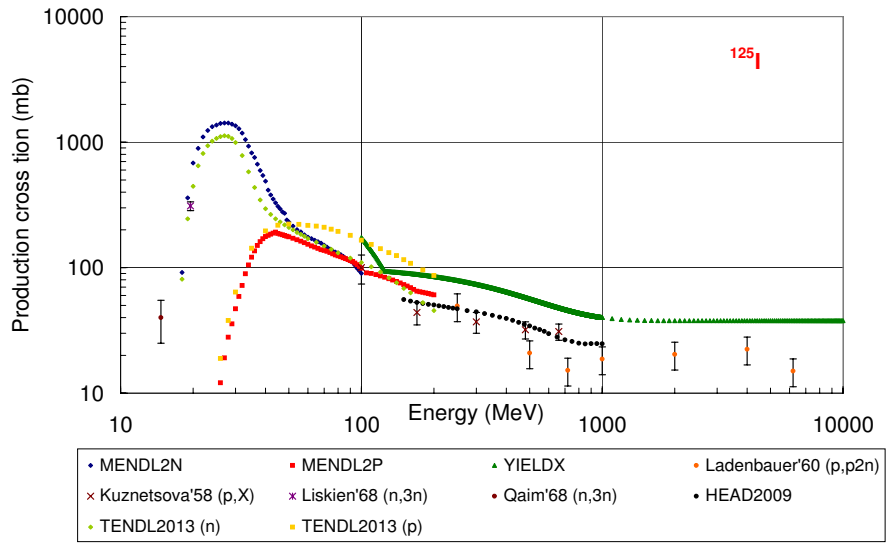}
 \caption{As figure~\ref{sigmai126}, for $^{125}$I.}
  \label{sigmai125}
\end{figure}

\begin{figure}
 \centering
 \includegraphics[width=12cm]{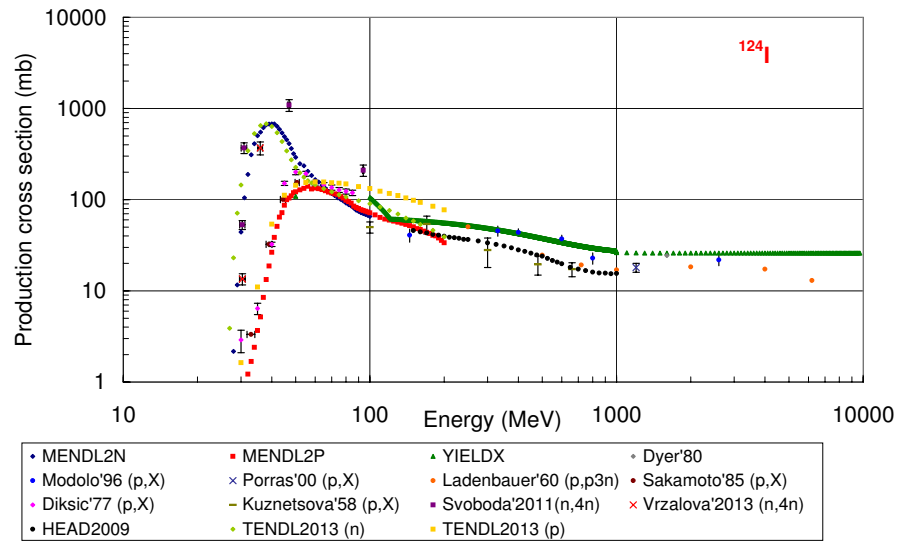}
 \caption{As figure~\ref{sigmai126}, for $^{124}$I.}
  \label{sigmai125}
\end{figure}

\begin{figure}
 \centering
 \includegraphics[width=12cm]{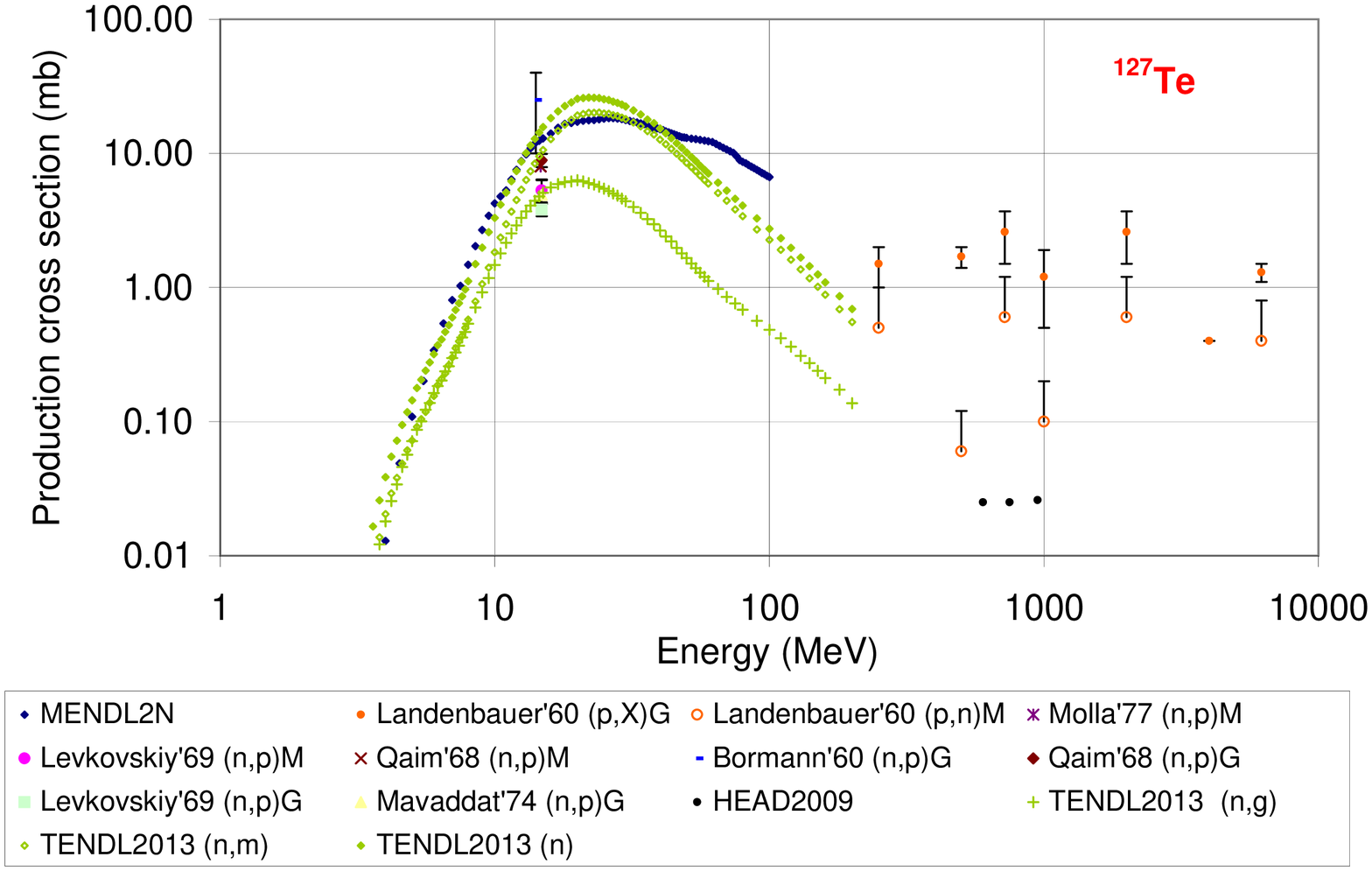}
 \caption{As figure~\ref{sigmai126}, for $^{127}$Te.}
  \label{sigmate127}
\end{figure}

\begin{figure}
 \centering
 \includegraphics[width=12cm]{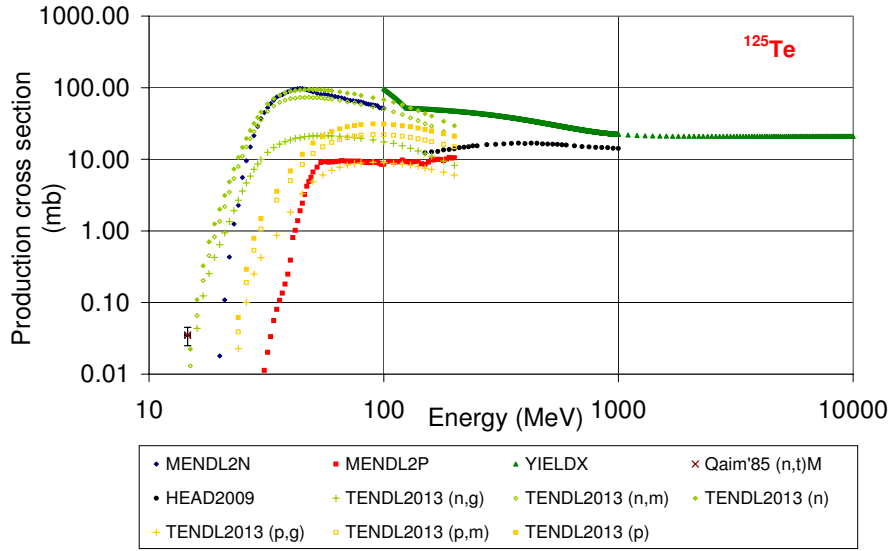}
 \caption{As figure~\ref{sigmai126}, for $^{125}$Te.}
  \label{sigmate125}
\end{figure}

\begin{figure}
 \centering
 \includegraphics[width=12cm]{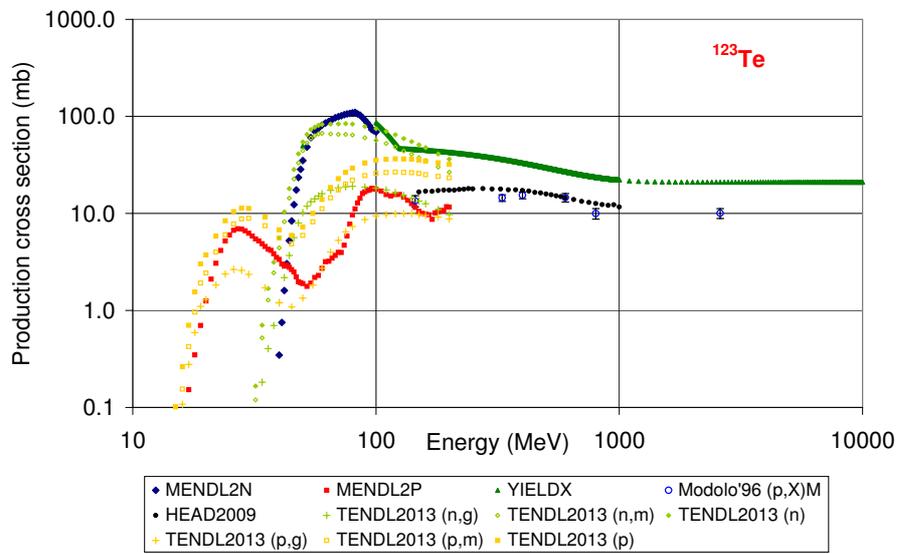}
 \caption{As figure~\ref{sigmai126}, for $^{123}$Te.}
  \label{sigmate123}
\end{figure}

\begin{figure}
 \centering
 \includegraphics[width=12cm]{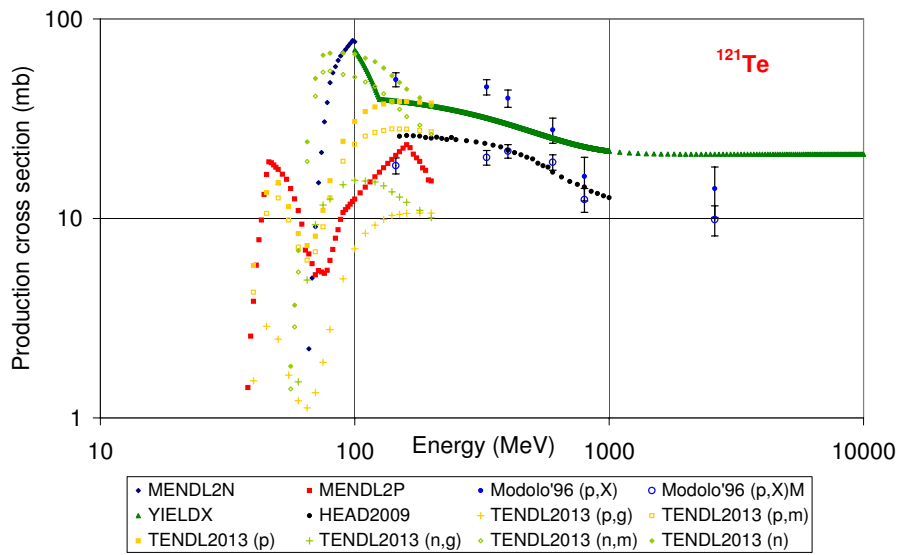}
 \caption{As figure~\ref{sigmai126}, for $^{121}$Te.}
  \label{sigmate121}
\end{figure}

\begin{figure}
 \centering
 \includegraphics[width=12cm]{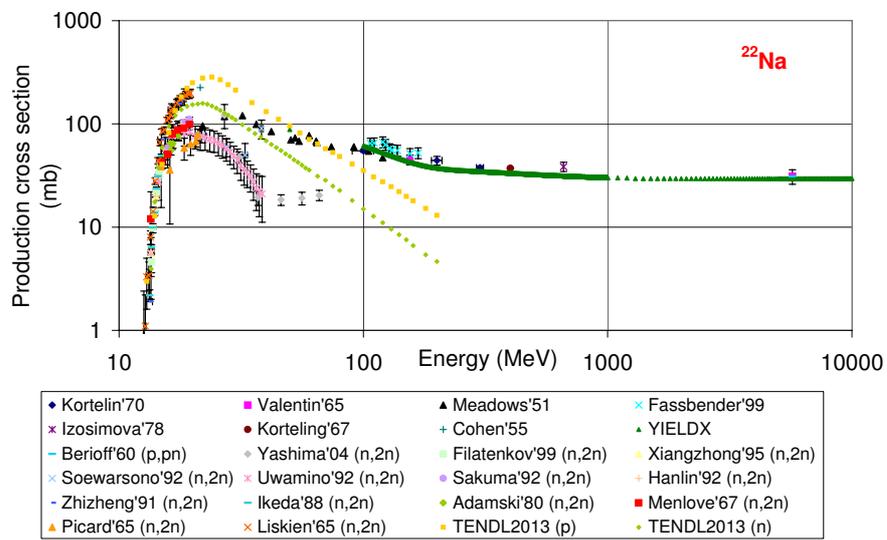}
 \caption{As figure~\ref{sigmai126}, for production of $^{22}$Na on $^{23}$Na.}
  \label{sigmana22}
\end{figure}

\end{document}